\documentstyle[12pt,epsfig]{article}
\def\bra{\langle}
\def\ket{\rangle}
\def\beq{\begin{equation}}
\def\eeq{\end{equation}}
\def\beqa{\begin{eqnarray}}
\def\eeqa{\end{eqnarray}}
\def\dspl{\displaystyle}
\def\para{{\scriptstyle\|}}
\def\tbst{\vrule width0pt height15pt depth10pt}

\begin{document}
\begin{flushright}
UH-511-982-01 \\
BELLE note \#419
\end{flushright}
\begin{center}
{\large Time-dependent Angular Analysis of $B$ Decays}

\vskip 0.4in
K. Abe\\
KEK, Tsukuba, Ibaraki, Japan 305-0801\\
and\\
M. Satpathy\\
Utkal University, Bhubaneswar, India\\
and\\
Hitoshi Yamamoto\\
University of Hawaii, Honolulu, HI 96822, USA

\vskip 0.4in
\parbox{5in}{
When both $B^0$ and $\bar B^0$ can decay to the same final state
composed of two vectors, the interference between them and those among
three polarization states result in intricate phenomena.
In this note we derive the time and angular
distributions for general $B \rightarrow V_a V_b$ processes
in a form convenient for actual analyses. 
We then apply them to specific examples and clarify the $CP$
violating parameters obtainable in the
$D^* \rho$ and $J/\psi K^*$ final states. 
The time distributions for the $D^* \pi$ final states are
also discussed.
}
\end{center}

\section{Angular dependence}

The essential parts of this and next section can be found in many
references~\cite{angtim}. Here, we attempt to describe central
concepts and derive critical expressions as simply as possible.

\subsection{Introduction}

We consider a two-body decay $0\to a+b$ in the rest frame of the
parent particle, where the spin state
$(J,M)$ of the parent particle and the helicities $\lambda_{a,b}$ of the
daughters are given.
The final state with a definite total angular momentum and
definite helicities can be constructed as follows:
In general, if $|\hat n \lambda\ket$ is a state with
total angular momentum $\lambda$ along the direction $\hat n$,
one can form a state with total angular momentum $|J,M\ket$
where the quantization axis is taken as the $z$ direction
(i.e. in the lab frame), as 
\beq
  |JM,\lambda\ket = 
  \int d\hat n\, D^{J*}_{M,\lambda}(\hat n)\,
  |\hat n \lambda\ket\,, 
  \label{eq:JMlam}
\eeq
with
\beq
  d\hat n = d\phi\, d\!\cos\theta\,,\quad
    D^{j}_{m, m'}(\hat n)  = D^{j}_{m, m'}(\phi,\theta,0)
\eeq
where $(\theta,\phi)$ is the polar coordinate of the direction
$\hat n$, and $D^{j*}_{m, m'}(\hat n)$ is the rotation function,
or the wave function of a top with total angular momentum
$|JM\ket$ and the component along $\hat n$ given by $\lambda$ which
is also a good quantum number. 

Suppose
$|\hat p \lambda_a\lambda_b\ket$ is the state in which
particle $a$ is moving in the $\hat p$ direction with helicity
$\lambda_a$ and particle $b$ is moving in the $-\hat p$ direction with helicity
$\lambda_b$:
\beq
   |\hat p \lambda_a\lambda_b\ket =
     |\hat p \lambda_a\ket |-\hat p\lambda_b\ket\,.
\eeq
Then, (\ref{eq:JMlam}) with the identification 
\beq
 \hat n = \hat p\,,\quad \lambda = \lambda_a - \lambda_b\,,
\eeq
gives the state with total angular momentum $|JM\ket$ and total helicity
$\lambda_a - \lambda_b$ along the direction of $a$:
\beq
  |JM,\lambda_a\lambda_b\ket = N
  \int d\hat p\, D^{J*}_{M\,\lambda_a-\lambda_b}(\hat p)\,
  |\hat p \lambda_a\lambda_b\ket\,,
 \label{eq:helstate}
\eeq
where $N$ is a normalization factor.
The ranges of the integration are
\beq
       -1\le\cos\theta\le1\,,\quad 0\le\phi\le 2 \pi\,.
\eeq
The possible values of the heclicities are constrained by
\beq
   |\lambda_a - \lambda_b| \le M\,,
  \label{eq:helconst}
\eeq
which arises since the orbital angular momentum cannot have
a component along the line of decay.
The construction (\ref{eq:helstate}) indicates that the amplitude
for particle $a$ to be in direction $\hat p$ is 
$D^{J*}_{M\,\lambda_a-\lambda_b}(\hat p)$.

Transformation of the state $|JM,\lambda_a\lambda_b\ket$ under parity is given
by~\cite{JW}
\beq
    P |JM,\lambda_a\lambda_b\ket = \pi_a \pi_b (-1)^{J-s_a-s_b}
      |JM,-\lambda_a-\lambda_b\ket\,,
 \label{eq:paritygen}
\eeq
where $s_{a,b}$ and 
$\pi_{a,b}$ are the spins and intrinsic parities of the daughter particles,
respectively.

\subsection{$B \to V_aV_b$, helicity basis}

In $B$ decays of the type $B\to V_a V_b$ ($V$: a vector), such as
$B^+\to \Psi K^{*+}$ and $\bar D^{*0}\rho^+$, we have
\beq
      J = M = 0,\quad s_a = s_b = 1,\quad \pi_a = \pi_b = -1\,.
 \label{eq:BVVvals}
\eeq
The constraint (\ref{eq:helconst}) with $M=0$ means $\lambda_a = \lambda_b$,
and thus there are three possible helicity states:
\beq
     (\lambda_a,\lambda_b) = (+1,+1),\quad (0,0), \quad{\rm or} \;(-1,-1)\,.
\eeq
Accordingly, the final state can be written as
\beq
   |\Psi_f\ket = \sum_\lambda H_\lambda\, |f_\lambda\,\ket \quad (\lambda = +1,0,-1)\,,
 \label{eq:fstate}
\eeq
where $H_i$ is the amplitude for each helicity state, and we have
defined 
\beq
   \begin{array}{rcl}
      |f_{+1}\ket &\equiv& |JM, +1 +1\ket\,,\\
      |f_{0}\ket &\equiv& |JM,  0 0 \ket\,,\\
      |f_{-1}\ket &\equiv& |JM, -1 -1\ket\,.
   \end{array}
\quad (J=M=0)
\eeq
In terms of decay amplitude, one can write
\beq
      H_\lambda = \bra f_\lambda|H_{\rm eff}|B\ket\,,
\eeq
where $H_{\rm eff}$ is the effective Hamiltonian responsible for the decay.

\begin{figure}
  \begin{center}
  \epsfig{file=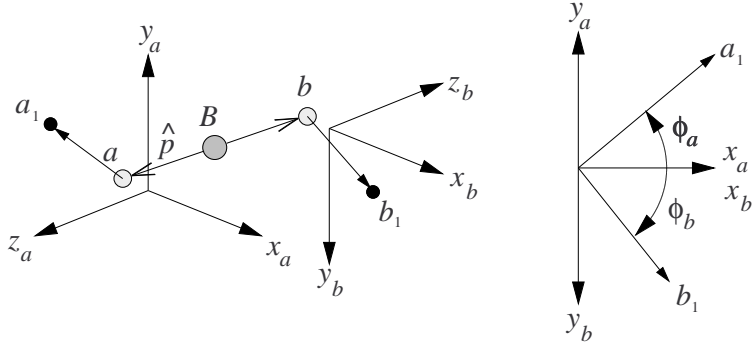, width=4.2in}
  \caption{Definition of angles for $B\to V_a V_b$ decay.}
  \label{fg:angledef}
  \end{center}
\end{figure}

When the daughter particles subsquently decay as
\beq
    a \to a_1 + a_2\,,\quad b\to b_1 + b_2\,,
\eeq
the construction (\ref{eq:helstate}) applies to each decay in its
rest frame. The decay amplitude for $a_1$ to be in direction
$(\theta_a,\phi_a)$ in the rest frame of $a$ and 
$b_1$ to be in direction $(\theta_b,\phi_b)$ in the rest frame of $b$ is 
then (up to an overall constant)
\beq
   A = \sum_m H_m \,
    D^{s_a *}_{m,\lambda_{a_1}-\lambda_{a_2}}(\phi_a,\theta_a,0)\,
    D^{s_b *}_{m,\lambda_{b_1}-\lambda_{b_2}}(\phi_b,\theta_b,0)\,.
\eeq
The $z$ axis in the rest frame of $a$ is taken to be in the direction
of $\hat p$, and that in the rest frame of $b$ is taken to be
in the direction of $-\hat p$; namely, each in the direction of
the motion of the parent particle in the $B$ frame. The definition of the
azimuthal angles amounts to defining the
phase convention for the helicity amplitudes $H_m$. To be specific,
we define that the $x$ directions in the two frames are the same
(see Figure~\ref{fg:angledef}).
Using
\beq
      D^j_{m, m'}(\alpha,\beta,\gamma)
   = e^{-im\alpha}d^j_{m, m'}(\beta) e^{-i m' \gamma}\,,
\eeq
the amplitude can be written as
\beq
 A = \sum_m H_m\;
   e^{i m\chi} d^{s_a}_{m,\lambda_{a_1}-\lambda_{a_2}}(\theta_a)
             d^{s_b}_{m,\lambda_{b_1}-\lambda_{b_2}}(\theta_b)\,,
 \label{eq:anggen}
\eeq
with
\beq
    \chi \equiv \phi_a + \phi_b
\eeq
being the azimuthal angle from $b_1$ to $a_1$ measured counter-clock-wise
looking down from the $a$ side.

\subsection{Transversity basis}

Using the values (\ref{eq:BVVvals}), the parity transformation
(\ref{eq:paritygen}) reads
\beq
      P|f_{+1}\ket = |f_{-1}\ket\,\quad
      P|f_{0}\ket = |f_0\ket\,\quad
      P|f_{-1}\ket = |f_{+1}\ket\,;
\eeq
namely, the helicity-basis states $|\pm1\ket$ are not
parity eigenstates. However, we can construct parity eigenstates
as
\beq
  \begin{array}{rclr}
  |f_\para\,\ket   &\equiv& \dspl
         {|f_{+1}\ket + |f_{-1}\ket\over\sqrt2} &\quad (P+) \\
  |f_\bot\ket &\equiv& \dspl
         {|f_{+1}\ket - |f_{-1}\ket\over\sqrt2} & (P-)
  \end{array}\,,\quad\hbox{ and }\quad
  |f_0\,\ket \quad (P+) \,.
\eeq
The final state (\ref{eq:fstate}) can then be written as
\beq
   |\Psi_f\,\ket = \sum_\lambda A_\lambda \, |f_\lambda\ket\qquad
   (\lambda = \para, 0, \bot)\,,
\eeq
with
\beq
     \begin{array}{rcl}
  A_\|   &\equiv& \dspl
         {H_+ + H_-\over\sqrt2} \\
  A_\bot &\equiv& \dspl
         {H_+ - H_-\over\sqrt2} 
  \end{array}\,,\quad\hbox{ and }\quad
  A_0 \equiv H_0 \,.
 \label{eq:tramps}
\eeq
This basis is called the transversity basis~\cite{Rosner}.

\begin{figure}
  \begin{center}
  \epsfig{file=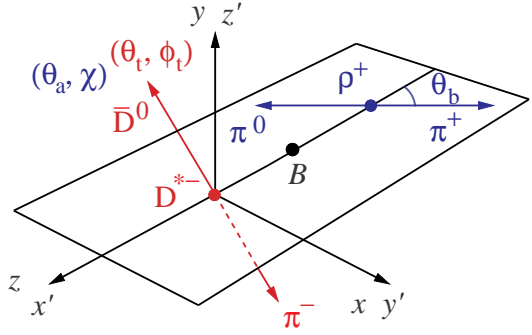, width=3in}
  \caption{Angles often used for the transversity basis are shown for
       $D^*\rho$ final state.}
  \label{fg:transbdrho}
  \end{center}
\end{figure}

An often-used set of angles for the transversity basis can be obtained as follows:
We note first that the angles $(\theta_a,\chi)$ defined in the previous
section is the polar coordinate of the $a_1$ direction in the
$a$ rest frame  where the $z$-direction
is taken to be opposite the direction of $b$ in that frame and
the $x$ direction is taken to be in the decay plane of 
$b\to b_1\, b_2$ such that $p_x(b_1)$ is positive. This defines a
right-handed coordinate system where the $y$ axis is perpendicular
to the decay plane. We now define a new right handed system by
\beq
      x' = z\,,\quad y'=x\,,\quad z'=y\,,
\eeq
where the $z'$-axis is now perpendicular to the $b$ decay plane.
Then, $(\theta_{tr},\theta_{tr})$ is defined as 
the polar coordinate of the $a_1$ in this new system. Namely,
$(\phi_{tr},\phi_{tr})$ and $(\theta_a,\chi)$ are related by
\beq
  \begin{array}{ccccccc}
    x' &=& \sin \theta_{tr} \cos \phi_{tr} &=& \cos \theta_a &=& z \\
    y' &=& \sin \theta_{tr} \sin \phi_{tr} &=& \sin \theta_a \cos \chi &=& x \\
    z' &=& \cos \theta_{tr} &=& \sin \theta_a \sin \chi &=& y
  \end{array}\,.
 \label{eq:trangs}
\eeq
These angles are shown for the case of $D^*\rho$ in Figure~\ref{fg:transbdrho}.
Note, however, that one could also use the angles $(\chi,\theta_a,\theta_b)$
for the transversity basis.

\subsubsection{$B \rightarrow \bar D^* \rho^+$ (helicity)}

A full angular analysis of this mode has been presented at 
conferences~\cite{CLEODstrho}, but has not been published.
Here, we consider the decay 
$B \to \bar D^{*} \rho^+$ which is followed by
\beq
   \bar D^{*}\to \bar D \pi\,,\quad \rho^+\to \pi^+\pi^0\,.
\eeq
We assign,
\beq
   a = \bar D^{*}\,,\quad 
   a_1 = \bar D\,,\quad a_2 = \pi\,,\quad
   b = \rho^+\,,\quad
   b_1 = \pi^+\,,\quad b_2 = \pi^0\,.
 \label{eq:dstrhoassign}
\eeq
The decays of $D^*$ and $\rho$ have only one helicity state:
\beq
    \lambda_{a_1}-\lambda_{a_2} = 0\,,\quad
    \lambda_{b_1}-\lambda_{b_2} = 0\,.
\eeq
Thus, the general amplitude form (\ref{eq:anggen}) becomes
\beq
  A = \sum_m H_m\;e^{i m\chi} d^1_{m,0}(\theta)
                                     d^1_{m,0}(\psi)\,,
  \label{eq:dstrho}
\eeq
where we have relabeled the polar angles
\beq
        \theta = \theta_a \;(D^*)\,,\quad \psi = \theta_b\; (\rho)\,.
\eeq
This can be rewritten as 
\begin{equation}
A = H_+ g_+ + H_0 g_0 + H_{-1} g_{-1}
\label{eq:amphelicity}
\end{equation}
where
\begin{eqnarray}
      g_{+1} &=& {1\over2 } e^{i\chi} \sin\theta \sin\psi \nonumber \\
      g_{0}  &=& \cos \theta \cos \psi \nonumber \\
      g_{-1} &=& {1\over2 } e^{-i\chi} \sin\theta \sin\psi . 
   \label{eq:gmhldstrho}
\end{eqnarray}
We have used
\beq
   d^1_{1,1}(\theta) = {1 + \cos \theta\over 2}\,,\quad
   d^1_{1,0}(\theta)  = -{\sin\theta\over \sqrt2}\,,\quad
   d^1_{1,-1}(\theta)  = {1 - \cos \theta\over 2}\,,
\eeq
together with
\beq
   d^j_{m,m'}(\theta) = (-)^{m-m'}d^j_{m',m}(\theta) = d^j_{-m',-m}(\theta)\,.
\eeq

The square of the amplitude is
\beqa
  |A|^2 &=&
   (\sum_m H_m\, g_m)^* (\sum_n H_n g_n)
        \nonumber \\
 &=& \sum_m |H_m|^2 |g_m|^2  \nonumber \\
  &&  + 2 \sum_{m<n}\Big( 
      \Re (H_m^* H_n) \,\Re (g^*_m g_n)
     -\Im (H_m^* H_n) \,\Im (g^*_m g_n) \Big)\,.
 \label{eq:ampsquare}
\eeqa
Using the explicit forms for $g_m$, the final
distribution is
\beqa
 \Gamma(\chi,\theta,\psi) &=&
   {9\over32\pi} \Big[ (|H_+|^2 + |H_-|^2) \sin^2\theta \sin^2\psi
    + 4|H_0|^2 \cos^2\theta \cos^2\psi  \nonumber \\
  && \hspace{-0.7in} +2 \Big\{ \Re(H_+ H_-^*) \cos {2\chi} - \Im(H_+ H_-^*)
    \sin{2\chi}\Big\}
         \sin^2 \theta \sin^2 \psi  \nonumber \\
  && \hspace{-0.7in} + \Big\{ \Re((H_+ + H_-) H_0^*) \cos \chi -\Im((H_+ - H_-) H_0^*) 
     \sin \chi \Big\} \sin {2\theta} \sin {2\psi} \Big]\,,
 \label{eq:angdisheldrho}
\eeqa
where the normalization factor is chosen such that
\beq
   \int_0^{2\pi} d\chi \int_{-1}^1 d\cos\theta \int_{-1}^1 d\cos\psi\;
   \Gamma(\chi,\theta,\psi) = 
  |H_+|^2 + |H_-|^2 + |H_0|^2\,.
\eeq

\subsubsection{$B \rightarrow D^* \rho$ (transversity)}

Here, we can transform the $D^*$ side to transversity angles, or we could
choose the $\rho$ side. We arbitrarily choose $D^*$ side.
We start from the amplitude (\ref{eq:amphelicity}) and apply the transformations
(\ref{eq:tramps}) and (\ref{eq:trangs}). We obtain
\beq
  \begin{array}{llll}
    g_\| &= {1\over\sqrt2}(g_+ + g_-) &= {1\over\sqrt2} \cos \chi \sin \theta \sin \psi 
     &=  {1\over\sqrt2} \sin \theta_{tr} \sin \phi_{tr} \sin \psi \\
    g_0 & &= \cos \theta \cos \psi & = \sin \theta_{tr} \cos \phi_{tr} \cos \psi \\
    g_\bot &= {1\over\sqrt2}(g_+ - g_-) &= {i\over\sqrt2} \sin \chi \sin \theta \sin \psi 
     &=  {i\over\sqrt2} \cos \theta_{tr} \sin \psi
  \end{array}\,,
   \label{eq:gmtrdstrho}
\eeq
to be used in
\beq
      A(\phi_{tr},\theta_{tr},\psi) = \sum_m A_m \, g_m (\phi_{tr},\theta_{tr},\psi)\quad
     (m = \para, 0, \bot)\,.
  \label{eq:dpiampt}
\eeq
Squaring this as before, the angular distribution becomes
\beqa
 {d^3 \Gamma(\phi_{tr},\theta_{tr},\psi)\over d\phi_{tr} d \cos \theta_{tr} d \cos \psi}
    &=& {9\over 32\pi} \bigg(
   |A_\| |^2 2 \sin^2\theta_{tr} \sin^2\phi_{tr} \sin^2\psi \nonumber\\
   &&\hspace{-1in} + |A_\bot |^2 2 \cos^2\theta_{tr} \sin^2\psi
   + |A_0|^2 4 \sin^2\theta_{tr} \cos^2\phi_{tr} \cos^2\psi \nonumber \\
   && \hspace{-1in} + \sqrt2 \Re(A_\|^* A_0) \sin^2\theta_{tr}
     \sin {2\phi_{tr}} \sin {2\psi}
   - \sqrt2 \Im(A_0^*A_\bot) \sin {2\theta_{tr}} \cos \phi_{tr} \sin {2\psi} \nonumber \\
   && \hspace{-1in} -2 \Im(A_\|^* A_\bot) \sin {2\theta_{tr}} \sin \phi_{tr} \sin^2\psi
\bigg)
  \label{eq:angtrdstrho}
\eeqa
Integrating this over $\phi_{tr}$ loses all interference effects among different
polarization states:
\beq
 {d^2 \Gamma(\phi_{tr},\theta_{tr},\psi)\over d \cos \theta_{tr} d \cos \psi} =
  {9\over16}  \bigg( |A_\||^2 \sin^2\theta_{tr} \sin^2\psi 
  + |A_0|^2 2 \sin^2\theta_{tr} \cos^2\psi
    + |A_\bot^2| 2 \cos^2\theta_{tr} \sin^2\psi \bigg)\,.
\eeq
At this point, we see that the even parity states ($A_\|$ and $A_0$) have $\sin^2\theta_{tr}$
distribution, and the odd parity state ($A_\bot$) has $\cos^2\theta_{tr}$ distribution.
Thus, plotting $\theta_{tr}$ distribution only can separate even and odd parity
components. On the other hand, both $A_\|$ and $A_\bot$ are associated with
$\sin^2\psi$, and thus $\psi$ distribution alone cannot separate different
parity components. Further integrating over $\psi$ gives
\beq 
 {d\Gamma(\phi_{tr},\theta_{tr},\psi)\over d \cos \theta_{tr}} =
  {3\over4}  \bigg( (|A_\||^2 + |A_0|^2) \sin^2\theta_{tr}
    + |A_\bot^2| 2 \cos^2\theta_{tr}\bigg)\,.
\eeq

\subsubsection{$B\to\Psi K^*$ (helicity)}

The time-independent analysis has been performed by many
experiments~\cite{CLEOpsikst}.
We assign 
\beq
   a = \Psi\,,\quad 
   a_1 = \ell^+\,,\quad a_2 = \ell^-\,,\quad
   b = K^*\,,\quad
   b_1 = K\,,\quad a_2 = \pi\,.\quad
\eeq
The decay $K^*\to K\pi$ has only one helicity state
$\lambda_{b_{1,2}} = 0$. On the other hand, the final state
of $\Psi\to\ell^+\ell^-$ can have multiple helicity states because
of the lepton spins. The actual helicity states, however, are
restricted to only two
due to the vector nature of the coupling that creates the
lepton pair:
\beq
  (\lambda_{\ell^+},\lambda_{\ell^-}) 
   = \Big(+{1\over2},-{1\over2}\Big)\;\hbox{ or }\;
     \Big(-{1\over2},+{1\over2}\Big)\,.
\eeq
We have thus,
\beq
    \lambda_{a_1}-\lambda_{a_2} = \pm1\,,\quad
    \lambda_{b_1}-\lambda_{b_2} = 0\,.
\eeq
The final angular distribution is given by incoherent sum of the distributions
for the two lepton helicity combinations:
\beq
   \Gamma(\chi,\theta,\psi) = |A^{(+1)}|^2 + |A^{(-1)}|^2\,,
\eeq
with
\beq
  A^{(\lambda)} = \sum_m H_m\;e^{i m\chi} d^1_{m\lambda}(\theta)
                                     d^1_{m0}(\psi)\,,
  \label{eq:aPsiKst}
\eeq
where we have relabeled the polar angles
\beq
        \theta = \theta_a \;(\Psi)\,,\quad \psi = \theta_b\; (K^*)\,.
\eeq
The amplitude (\ref{eq:aPsiKst}) has the form
\beq
    A^{(\lambda)} = \sum_m H_m\, g^{(\lambda)}_m\,,
  \label{eq:aPsiKstshort}
\eeq
where
\beq
   \begin{array}{rcl}
      g^{(+1)}_{+1} &=& -{1\over2\sqrt2} (1+\cos \theta) e^{i\chi} \sin \psi \\
      g^{(+1)}_{0}  &=&  {1\over\sqrt2} \sin \theta \cos \psi \\
      g^{(+1)}_{-1} &=&  {1\over2\sqrt2} (1-\cos \theta) e^{-i\chi} \sin \psi 
   \end{array}\,,\quad
   \begin{array}{rc@{\dspl}l}
      g^{(-1)}_{+1} &=& -{1\over2\sqrt2} (1-\cos \theta) e^{i\chi} \sin \psi \\
      g^{(-1)}_{0}  &=& -{1\over\sqrt2} \sin \theta \cos \psi \\
      g^{(-1)}_{-1} &=&  {1\over2\sqrt2} (1+\cos \theta) e^{-i\chi} \sin \psi 
   \end{array}\,.
\eeq

The square of the amplitude (\ref{eq:aPsiKstshort}) is
\beqa
  |A^{(\lambda)}|^2 &=&
   (\sum_m H_m\, g^{(\lambda)}_m)^* (\sum_n H_, g^{(\lambda)}_n)
        \nonumber \\
 &=& \sum_m |H_m|^2 |g^{(\lambda)}_m|^2  \nonumber \\
  &&  + 2 \sum_{m<n}\Big( 
      \Re (H_m^* H_n) \,\Re (g^{(\lambda)*}_m g^{(\lambda)}_n)
     -\Im (H_m^* H_n) \,\Im (g^{(\lambda)*}_m g^{(\lambda)}_n) \Big)\,.
 \label{eq:asquare}
\eeqa
Using the explicit forms for $g^{(\lambda)}_m$, the final
distribution is
\beqa
 \Gamma(\chi,\theta,\psi) &=&
   {9\over64\pi} \Big[ (|H_+|^2 + |H_-|^2)(1+\cos \theta^2)\sin^2\psi
    + |H_0|^2\, 4 \sin^2\theta \cos^2\psi  \nonumber \\
  && \hspace{-0.7in} - 2 \Big\{ \Re(H_+^* H_-) \cos {2\chi} 
      + \Im(H_+^* H_-) \sin {2\chi}\Big\}
         \sin^2\theta \sin^2\psi  \nonumber \\
  && \hspace{-0.7in} - \Big\{ \Re((H_+ + H_-)^* H_0) \cos \chi 
      +\Im((H_+ - H_-)^* H_0) \sin \chi \Big\}
   \sin {2\theta} \sin {2\psi} \Big]\,,
 \label{eq:angdishel}
\eeqa
where the normalization factor is chosen such that
\beq
   \int_0^{2\pi} d\chi \int_{-1}^1 d\cos\theta \int_{-1}^1 d\cos\psi\;
   \Gamma(\chi,\theta,\psi) = 
  |H_+|^2 + |H_-|^2 + |H_0|^2\,.
\eeq

\subsubsection{$B\to\Psi K^*$ (transversity)}

One can use the relations (\ref{eq:tramps}) and (\ref{eq:trangs})
directly in the angular distribution (\ref{eq:angdishel}) to
obtain
\beqa
  \Gamma(\phi_{tr},\theta_{tr},\psi) &=&
    {9\over32\pi}\Big[ |A_\| |^2 (1-\sin^2\theta_{tr} \sin^2\phi_{tr}) \sin^2\psi
      \nonumber \\
    &&\hspace{-0.7in}  + |A_0|^2 2(1-\sin^2\theta_{tr} \cos^2\phi_{tr}) \cos^2\psi
     + |A_\bot|^2 \sin \theta_{tr}^2 \sin \psi^2 \nonumber \\
   && \hspace{-0.7in}- \Re(A_\|^* A_0) {1\over\sqrt2} 
     \sin \theta_{tr}^2 \sin {2\phi_{tr}} \sin{2\psi}
      + \Im(A_0^* A_\bot) {1\over\sqrt2} \sin {2\theta_{tr}} \cos \phi_{tr} \sin {2\psi}
      \nonumber \\
    &&\hspace{-0.7in} + \Im(A_\|^* A_\bot) \sin {2\theta_{tr}} \sin \phi_{tr} \sin^2\psi\Big] \,,
  \label{eq:angdistr}
\eeqa
which is normalized as
\beq
   \int_0^{2\pi} d\phi_{tr} \int_{-1}^1 d\cos\theta_{tr} \int_{-1}^1 d\cos\psi\;
   \Gamma(\phi_{tr},\theta_{tr},\psi) = 
  |A_\||^2 + |A_\bot|^2 + |A_0|^2\,.
\eeq

The transformation of angles can also be done at amplitude level.
With the substitution of ampltudes (\ref{eq:tramps}), 
the amplitude for a given lepton total helicity $\lambda$ becomes
\beq
   A^{(\lambda)} = \sum_m A_m\, g^{(\lambda)}_m\,,
 \quad (m = \para, 0, \bot)\,,
\eeq
with
\beq
   \begin{array}{rcl}
      g^{(+1)}_{\|} &=& -{1\over2} (\cos \chi \cos \theta + i \sin \chi) \sin \psi \\
      g^{(+1)}_{0}  &=&  {1\over\sqrt2} \sin \theta \cos \psi \\
      g^{(+1)}_\bot &=& -{1\over2} (\cos \chi + i \cos \theta \sin \chi) \sin \psi 
   \end{array}\,,
\eeq
\beq
   \begin{array}{rc@{\dspl}l}
      g^{(-1)}_{\|} &=&  {1\over2} (\cos \chi \cos \theta - i \sin \chi) \sin \psi \\
      g^{(-1)}_{0}  &=& -{1\over\sqrt2} \sin \theta \cos \psi \\
      g^{(-1)}_\bot &=& -{1\over2} (\cos \chi - i \cos \theta \sin \chi) \sin \psi
   \end{array}\,.
\eeq
In order to apply the transformation from $(\theta,\chi)$ to
$(\theta_{tr},\phi_{tr})$, it is easier to multiply an overall phase factor which
does not affect the final angular distribution. We take
\beqa
  \xi^{(+1)} &=&
  {-\cos \chi + i \cos \theta \sin \chi \over \sin \theta_{tr}}\quad{\rm for}\quad g^{(+1)}_m\,,
  \nonumber \\ \xi^{(-1)} &=&
  { \cos \chi + i \cos \theta \sin \chi \over \sin \theta_{tr}}\quad{\rm for}\quad g^{(-1)}_m\,.
\eeqa
It is easy to see that these factors are indeed pure phases using the
relations (\ref{eq:trangs}):
\beqa
  \sin^2\theta_{tr} &=& \underbrace{\sin^2\theta_{tr} \cos^2\phi_{tr}}_{\dspl \cos^2\theta}
    + \underbrace{\sin^2\theta_{tr} \sin^2\phi_{tr}}_{\dspl \sin^2\theta \cos^2\chi}
    = \cos \theta^2 +  \hspace{-0.3in}
    \underbrace{\sin^2\theta \cos^2\chi}_{\dspl (1-\cos^2\theta)(1-\sin^2\chi)}
   \hspace{-0.3in} \nonumber \\
   &=&1-\sin^2\chi+\cos^2\theta \sin^2\chi = \cos^2\chi 
    + \cos^2\theta \sin^2\chi\,,
\eeqa
\beq
   \to\quad \left|  {-\cos \chi + i \cos \theta \sin \chi \over \sin \theta_{tr}} \right|^2 =
       \left|  {\cos \chi + i \cos \theta \sin \chi \over \sin \theta_{tr}} \right|^2 = 1\,.
\eeq
Multiplying $\xi^{(+1)}$ to $g^{(+1)}_{\|}$, we have
\beq
   g^{(+1)}_{\|} \to g^{(+1)}_{\|} \xi^{(+1)} = 
   -{1\over2} (\cos \chi \cos \theta + i \sin \chi) \sin \psi 
   {-\cos \chi + i \cos \theta \sin \chi \over \sin \theta_{tr}}\,.
\eeq
Using
\beqa
  (\cos \chi \cos \theta + i \sin \chi)(-\cos \chi + i \cos \theta \sin \chi) &=&
   - \cos \theta -i \sin \theta^2 \sin \chi \cos \chi \nonumber \\ &&\hspace{-1in} 
   = -\sin \theta_{tr} (\cos \phi_{tr} + i \cos \theta_{tr} \sin \phi_{tr})\,,
\eeqa
the phase-rotated $g^{(+1)}_{\|}$ is then
\beq
   g^{(+1)}_{\|} = {1\over2} (\cos \phi_{tr} + i \cos \theta_{tr} \sin \phi_{tr}) \sin \psi\,.
\eeq
Other functions are similarly obtained:
\beq
 \begin{array}{rcl}
   g^{(+1)}_{\|} &=&  {1\over2} (\cos \phi_{tr} + i \cos \theta_{tr} \sin \phi_{tr}) \sin \psi \\
   g^{(+1)}_{0}  &=&  {1\over\sqrt2} (-\sin \phi_{tr} + i \cos \theta_{tr} \cos \phi_{tr}) \cos \psi\\
   g^{(+1)}_\bot &=&  {1\over2} \sin \theta_{tr} \sin \psi 
 \end{array}\,,
\eeq
\beq
   \begin{array}{rc@{\dspl}l}
   g^{(-1)}_{\|} &=&  {1\over2} (\cos \phi_{tr} - i \cos \theta_{tr} \sin \phi_{tr}) \sin \psi \\
   g^{(-1)}_{0}  &=&  {1\over\sqrt2} (-\sin \phi_{tr} - i \cos \theta_{tr} \cos \phi_{tr}) \cos \psi\\
   g^{(-1)}_\bot &=& -{1\over2} \sin \theta_{tr} \sin \psi 
   \end{array}\,.
\eeq
These functions gives
\beq
 \begin{array}{@{\tbst}c}
  \sum_\lambda |g^{(\lambda)}_\||^2 
          = {1\over2}(1-\sin^2\theta_{tr} \sin^2\phi_{tr}) \sin^2\psi \,,\\
  \sum_\lambda |g^{(\lambda)}_0|^2 = (1-\sin^2\theta_{tr} \cos^2\phi_{tr}) \cos^2\psi \,,\quad
  \sum_\lambda |g^{(\lambda)}_\bot|^2 = {1\over2} \sin^2\theta_{tr} \sin^2\psi\,, \\
  \sum_\lambda \Re(g^{(\lambda)*}_\| g^{(\lambda)}_0) =
      -{1\over4\sqrt2} \sin^2\theta_{tr} \sin {2\phi_{tr}} \sin {2\psi} \,,\quad
  \sum_\lambda \Im(g^{(\lambda)*}_\| g^{(\lambda)}_0) = 0\,, \\
  \sum_\lambda \Re(g^{(\lambda)*}_\bot g^{(\lambda)}_0) = 0 \,,\quad
  \sum_\lambda \Im(g^{(\lambda)*}_\bot g^{(\lambda)}_0) = 
        {1\over4\sqrt2} \sin {2\theta_{tr}} \cos \phi_{tr} \sin {2\psi} \\
  \sum_\lambda \Re(g^{(\lambda)*}_\| g^{(\lambda)}_\bot) = 0 \,,\quad
  \sum_\lambda \Im(g^{(\lambda)*}_\| g^{(\lambda)}_\bot) = 
        -{1\over4} \sin {2\theta_{tr}} \sin \phi_{tr} \sin^2\psi \,,
 \end{array}
\eeq
which immediately leads to (\ref{eq:angdistr}) through (\ref{eq:asquare})
where $H_m$ are relaced by $A_m$.
Note that three of the combinations are zero; this arises from cancellations
between the two lepton helicities $\lambda = \pm1$.

\subsection{Charge conjugate decays}

For the charge conjugate decays ($\bar B$ decays), the rule for the
definitions of angles is to start
from the corresponding $B$ decay, exchange paritcles and antiparticles,
and then apply the definition of angles as if the daughter particles
were the original particles from the $B$ decay. For example,
for the decays corresponding to assignment (\ref{eq:dstrhoassign}) for
$B\to \bar D^*\rho^+$, the particles in the decay $\bar B\to D^*\rho^-$
are assined as
\beq
   a = D^{*}\,,\quad 
   a_1 = D\,,\quad a_2 = \pi\,,\quad
   b = \rho^-\,,\quad
   b_1 = \pi^-\,,\quad b_2 = \pi^0\,.
\eeq
and the angles $(\theta,\chi,\psi)$ are defined in the same way in terms of
$a_{1,2}$ and $b_{1,2}$. In particular, the angle $\chi$ is
the azimuthal angle from $b_1$ to $a_1$ measured counter-clock-wise
looking down from the $a$ side.

With this definition, the angular distribution is given by (\ref{eq:anggen})
with replacement
\beq
      H_\lambda \to \bar H_\lambda\,,
   \label{eq:helcp}
\eeq
with
\beq
     \bar H_\lambda \equiv \bra \bar f_\lambda|H_{\rm eff}|\bar B\ket\,.
\eeq
When $CP$ is conserved in decay, then we can take (see Appendix)
\beq
     \bar H_\lambda = H_{-\lambda} \quad (CP)\,,
 \label{eq:cphel}
\eeq
which holds to all orders in perturbation theory. 
In the literature, one sometimes
encounters a $CPT$ relation $\bar H_\lambda = H_{-\lambda}^*$ which is
correct only to first order in perturbation theory. This $CPT$ relation is thus not
applicable to the decays of concern where
the strong phases play inmportant role, since those phases
are higher order effects.
In terms of tranversity amplitudes, the $CP$ relation (\ref{eq:helcp}) reads
\beq
    \bar A_\| = A_\|\,,\quad
    \bar A_\bot = - A_\bot\,,\quad
    \bar A_0 = A_0\,.\quad (CP)
  \label{eq:cptrans}
\eeq

Inspecting the expressions for the angular 
distribution, one notes that moving from $B$ decay to
$\bar B$ decay according to (\ref{eq:cphel}) or (\ref{eq:cptrans})
corresponds to changing $\chi$ to $-\chi$ 
for the helicity formulation,
and $\theta_{tr}\to \pi-\theta_{tr}$ for the transversity formulation. 
These are nothing but the parity transformation
(or equivalantly the mirror inversion) of the configuration.
Namely, if one exchanges particles and antiparticles and take mirror inversion,
then the resulting angular distribution is the correct one, which is to say
that $CP$ is conserved.

\section{Time-dependence}

In this section, we will develop a formalism suited for neutral $B$ decays to
final states that are not $CP$ eigenstates. In later sections, it will
be applied to $D^{*+}\pi^-$ final state as well as each of the three polarization
states of $D^{*+}\rho^-$ or $\Psi K^*$.

First, let us recall the time evolution of pure $B^0$ and $\bar B^0$ states.
Assuming $CPT$, the physical states $B_a$ and $B_b$ can be written as
\beq
  \begin{array}{rll}
    B_a &= p B^0 + q \bar B^0 &\quad (m_a,\gamma_a) \\
    B_b &= p B^0 - q \bar B^0 &\quad (m_b,\gamma_b)
  \end{array}\,,
   \label{eq:babpq}
\eeq
where $m_{a,b}$ and $\gamma_{a,b}$ are the masses and decay rates of the
corresponding physical states. 
Theoretically and experimentally, $|p|=|q|$ within error of order 1\%.
Here, we assume $|p|=|q|$ which makes $p/q$ a pure phase factor. The lowest
order estimation gives (see Appendix)
\beq
  {p\over q} =  -{V_{td}^* V_{tb} \over V_{td} V_{tb}^*}\,.
 \label{eq:povq}
\eeq
which corresponds to the choice of the $CP$ phase of the neutral $B$ meson
given by
\beq
   CP|B^0\ket = \eta_B|\bar B^0\ket\,,\quad
  CP|\bar B^0\ket = \eta_B^* |B^0\ket\,,\quad{\rm with}\quad \eta_B = 1.
\eeq
The above value of $p/q$ is for the case $B_a$ is heavier than $B_b$:
\beq
    m_a > m_b\,.
\eeq

The physical states evolve as
\beq
    B_a \to B_a e^{-im_a t - {\gamma_a\over2}t} \,,\quad
    B_b \to B_b e^{-im_b t - {\gamma_b\over2}t} \,.
\eeq
Hereafter, we will assume that the decay rates of the two physical states are
the same
\beq
  \gamma_a = \gamma_b \equiv \gamma\,.
\eeq
then, the factor $e^{-{\gamma\over2}t}$ decouples from all amplitudes, which
we will drop for now and restore it at the end. We also separate an overall
phase factor $\exp(-i{m_a+m_b\over2}t)$ and discard it since such overall phase
factors do not affect measurable quantities. Then the evolutions of 
$B_{a,b}$ can be simplified as
\beq
    B_a \to B_a e^{-i{\delta m\over2}t}\,,\quad B_b \to B_b e^{i{\delta m\over2}t}\,
   \qquad (\times e^{-{\gamma\over2}t})\,,
\eeq
with
\beq
   \delta m \equiv m_a - m_b  > 0\,.
\eeq
Then, the time evolutions of pure $B^0$ and $\bar B^0$ can be obtained by
solving (\ref{eq:babpq}) for $B^0$ and $\bar B^0$ and then applying 
the time evolutions above:
\beq
   B^0 = {B_a + B_b\over 2p} \to {1\over2p}
     (\hspace{-0.2in} \underbrace{B_a}_{\dspl p B^0 + q \bar B^0}
      \hspace{-0.3in} e^{-i{\delta m\over2}t} 
   +  \hspace{-0.2in}\underbrace{B_b}_{\dspl p B^0 - q \bar B^0} 
      \hspace{-0.3in} e^{i{\delta m\over2}t})  
   = B^0 \cos{\delta m t\over2} - {q\over p}\bar B^0 i\sin{\delta m t\over2}\,.
\eeq
The time evolution of $\bar B$ is similarly obtained. Restoring the decay
factor $e^{-{\gamma\over2}t}$,
\beqa
     B^0 &\to& e^{-{\gamma\over2}t} \left(
     B^0 \cos{\delta m\, t\over2} - {q\over p}\bar B^0 i\sin{\delta m\, t\over2}
     \right) \,,\nonumber \\
 \bar B^0 &\to& e^{-{\gamma\over2}t} \left(
     \bar B^0 \cos{\delta m\, t\over2} - {p\over q} B^0 i\sin{\delta m\, t\over2}
     \right)\,.
  \label{eq:bbbarevol}
\eeqa

We now consider the decay amplitudes for a pure $B^0$ or $\bar B^0$ state
at $t=0$ to decay to a final state $f$ or its charge conjugate state
$\bar f$ at time $t$. 
The final state could be $D^{(*)-}\pi^+$ or any given polarization
state of $D^{*-}\rho^+$ or $\Psi K^{*0}$.
Define four instantaneous dedcay amplitudes by
\beq
  \begin{array}{rcl}
          a &\equiv& Amp(B^0\to f) \\
     \bar a &\equiv& Amp(\bar B^0\to \bar f) \\
          b &\equiv& Amp(B^0\to \bar f) \\
     \bar b &\equiv& Amp(\bar B^0\to f) 
  \end{array}\,,
  \label{eq:fouramps}
\eeq
For $f = D^{(*)-}\pi^+$, for example, $a$ and $\bar a$ are the favored amplitudes and
$b$ and $\bar b$ are the suppressed amplitudes. Then, (\ref{eq:bbbarevol})
gives
\beq
  \begin{array}{rcl@{\tbst}l}
    A_{B^0\to f}(t) &=& e^{-{\gamma\over2}t} \left(
     a \cos{\delta m\, t\over2} - {q\over p}\bar b\, i\sin{\delta m\, t\over2}
     \right) &=
      e^{-{\gamma\over2}t} a \left(
     \cos{\delta m\, t\over2} - \rho\, i\sin{\delta m\, t\over2}
     \right) \\
    A_{\bar B^0\to \bar f}(t) &=& e^{-{\gamma\over2}t} \left(
     \bar a \cos{\delta m\, t\over2} - {p\over q} b\, i\sin{\delta m\, t\over2}
     \right) &=
      e^{-{\gamma\over2}t} \bar a\left(
     \cos{\delta m\, t\over2} - \bar\rho\, i\sin{\delta m\, t\over2}
     \right) \\
    A_{B^0\to \bar f}(t) &=& e^{-{\gamma\over2}t} \left(
     b \cos{\delta m\, t\over2} - {q\over p}\bar a\, i\sin{\delta m\, t\over2}
     \right) &=
      e^{-{\gamma\over2}t} \bar a\left(
      \bar \rho \cos{\delta m\, t\over2} - i\sin{\delta m\, t\over2}
     \right)  \\
    A_{\bar B^0\to f}(t) &=& e^{-{\gamma\over2}t} \left(
     \bar b \cos{\delta m\, t\over2} - {p\over q}a\, i\sin{\delta m\, t\over2}
     \right) &=
      e^{-{\gamma\over2}t} a \left(
     \rho\cos{\delta m\, t\over2} - i\sin{\delta m\, t\over2}
     \right) 
  \end{array}
  \label{eq:timeamp4}
\eeq
with
\beq
  \rho\equiv {q\, \bar b\over p\, a}\,,\quad
  \bar\rho\equiv {p\, b\over q\, \bar a}\,.
  \label{eq:rhodef}
\eeq
For the bottom two amplitudes (the `suppressed' decays), we have ignored
overall phase
factors $p/q$ and $q/p$ for the second equalities. 

At this point, we can see the relation between the `suppressed' and `favored'
modes; namely, up to an overall phase, $\delta m t\to \delta m t +\pi$
transforms $A_{B^0\to f}(t)$ to $A_{\bar B^0\to f}(t)$ and
$A_{B^0\to f}(t)$ to $A_{\bar B^0\to f}(t)$. Equivalently, in the expressions of
decay rates,
\beq
      (\cos\delta mt, \sin\delta mt) \leftrightarrow
      (-\cos\delta mt, -\sin\delta mt) 
  \label{eq:supfavtransf}
\eeq
transforms between a suppressed mode and its favoed mode with the
same final state. Also, 
$p/q$ is the complex conjugate of $q/p$ (within the approximation
that $|p| = |q|$), and
as we will see more explicity later,
the weak phase of $b/ \bar a$ is the complex conjugate of that
of $\bar b/a$ with the rest being the `strong phase' which is
common to both. Thus, 
\beq
    (\hbox{weak phase}) \leftrightarrow -(\hbox{weak phase})
   \label{eq:BBbartransf}
\eeq
keeping the strong phase the same
transforms between a $B^0$ decay and the corresponding $\bar B^0$
decay (both `suppressed' or both `favored') apart from
the difference between $a$ and $\bar a$. Often $|a|$ and $|\bar a|$ are 
the same and if so the above transformation is exact in the
decay rates. When we extend the above time-dependent amplitudes to
include interferences between polarizations, the rule
between the same final state (\ref{eq:supfavtransf}) still
holds, but the relation between $B^0$ and $\bar B^0$ (\ref{eq:BBbartransf})
does not hold in the helicity basis. We will see, however, that
it holds in the transversity basis.

The time dependent
rates are obtained by squaring 
(\ref{eq:timeamp4}):
\beq
 \begin{array}{rc@{\tbst}l}
    \Gamma_{B^0\to f}(t) &=& |a|^2
      {e^{-\gamma t}\over2} \left[ (1+|\rho|^2) + (1-|\rho|^2)\cos\delta mt 
                         +  2\Im\rho \,\sin\delta mt \right] \\
    \Gamma_{\bar B^0\to \bar f}(t) &=& |\bar a|^2
      {e^{-\gamma t}\over2} \left[ (1+|\bar\rho|^2) + (1-|\bar\rho|^2)\cos\delta mt 
                         +  2\Im\bar\rho \,\sin\delta mt \right] \\
    \Gamma_{B^0\to f}(t) &=& |\bar a|^2
      {e^{-\gamma t}\over2} \left[ (1+|\bar\rho|^2) - (1-|\bar\rho|^2)\cos\delta mt 
                         -  2\Im\bar\rho \,\sin\delta mt \right] \\
    \Gamma_{\bar B^0\to f}(t) &=& |a|^2
      {e^{-\gamma t}\over2} \left[ (1+|\rho|^2) - (1-|\rho|^2)\cos\delta mt 
                         -  2\Im\rho \,\sin\delta mt \right]
 \end{array}\,.
  \label{eq:fourrates}
\eeq
In deriving this formula, we have assumed $CPT$ in the mixing and that
$\gamma_a = \gamma_b$. Otherwise, it is general; in particular, there
could be direct $CP$ violations in any of the decay amplitudes such as
$|a|\not= |\bar a|$ etc.

On $\Upsilon(4S)$, one would flavor-tag the other side by, say, a lepton.
If the tag side decays to $\ell^-$ at proper time $t_{tag}$, the quantum
correlation is such that the signal side is pure $B^0$ at the same proper
time $t_{sig} = t_{tag}$ and proceed to evolve as usual from that time
on. Thus, for $t_{sig}>t_{tag}$, the decay distribution is given simply
by the replacement 
\beq
  t \to \Delta t \equiv t_{sig} - t_{tag}\,.
\eeq
For $t_{sig}<t_{tag}$, all that is needed is to put absolute value on $\Delta t$
of the decay factor $e^{-\gamma \Delta t}$. Namely, (\ref{eq:fourrates})
becomes the distributions on $\Upsilon(4S)$ with the replacement
\beq
   \gamma t \to \gamma |\Delta t|\quad{\rm and}\quad
   \delta m t \to \delta m \Delta t\,.
\eeq
Explicitly,
\beq
 \begin{array}{rc@{\tbst}l}
    \Gamma_{\ell^-,f}(\Delta t) &=& |a|^2
      {e^{-\gamma |\Delta t|}\over2} 
                  \left[ (1+|\rho|^2) + (1-|\rho|^2)\cos\delta m\Delta t 
                         +  2\Im\rho \,\sin\delta m\Delta t \right] \\
    \Gamma_{\ell^+,\bar f}(\Delta t) &=& |\bar a|^2
      {e^{-\gamma |\Delta t|}\over2} \left[ (1+|\bar\rho|^2) 
                     + (1-|\bar\rho|^2)\cos\delta m\Delta t 
                         +  2\Im\bar\rho \,\sin\delta m\Delta t \right] \\
    \Gamma_{\ell^-,f}(\Delta t) &=& |\bar a|^2
      {e^{-\gamma |\Delta t|}\over2} \left[ (1+|\bar\rho|^2) 
                    - (1-|\bar\rho|^2)\cos\delta m\Delta t 
                         -  2\Im\bar\rho \,\sin\delta m\Delta t \right] \\
    \Gamma_{\ell^+, f}(\Delta t) &=& |a|^2
      {e^{-\gamma |\Delta t|}\over2} \left[ (1+|\rho|^2) 
                        - (1-|\rho|^2)\cos\delta m\Delta t 
                         -  2\Im\rho \,\sin\delta m\Delta t \right]
 \end{array}\,,
  \label{eq:fourrates4s}
\eeq
where $\Gamma_{\ell^-,f}(\Delta t)$ denotes the decay rate for one side decaying to
a final state $f$ while the opposite side is tagged by a negative lepton
(or tagged as $\bar B^0$ by any other method), etc.

\subsection{$B^0  \to D^{(*)-} \pi^+$}

\begin{figure}
  \begin{center}
    \epsfig{file=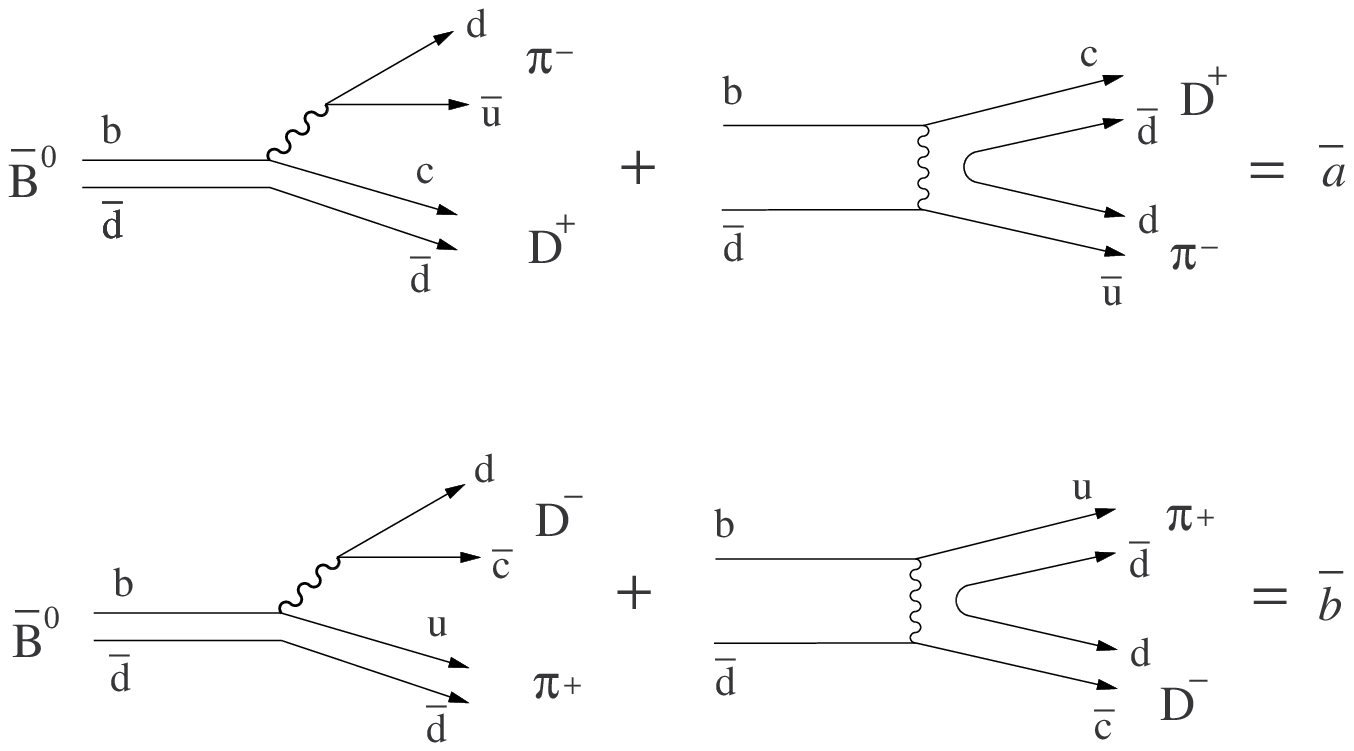,width=4.2in}
   \caption{Diagrams for $\bar B^0\to D^\pm\pi^\mp$.}
   \label{fg:bdpiamps}
  \end{center}
\end{figure}

Earlier studies of this mode can be found in Ref.~\cite{Dstpi}.
Diagrams for $\bar B^0 \to D^\mp \pi^\pm$ are
shown in Figure~\ref{fg:bdpiamps}. 
In addition to dominant tree
diagrams, annihilation diagrams may have non-negligible
contribution. Also, there may be final-state rescattering
$\bar D^{(*)0}\pi^0\to D^{(*)-} \pi^+$.
The $CKM$ factor of these processes, however, 
is the same as that of the corresponding tree diagram for the same
final state, and thus it does not affect the following formulation.
Penguins should result in even number of charms; thus,
penguins do not contribute. 

With the definitions
$f\equiv D^-\pi^+$ and $\bar f\equiv D^+\pi^-$,
the four amplitudes of (\ref{eq:fouramps}) can be written as
\beq
  \begin{array}{@{\tbst}rcl}
          a &\equiv& Amp(B^0\to D^-\pi^+) = \lambda_c^* F_c \\
     \bar a &\equiv& Amp(\bar B^0\to D^+\pi^-) =\lambda_c \bar F_c \\
          b &\equiv& Amp(B^0\to D^+\pi^-) = \lambda_u^* F_u\\
     \bar b &\equiv& Amp(\bar B^0\to D^-\pi^+) = \lambda_u \bar F_u 
  \end{array}\quad{\rm with}\quad
  \begin{array}{rcl}
     \lambda_c \equiv V_{cb} V_{ud}^* \\
     \lambda_u \equiv V_{ub} V_{cd}^*
  \end{array}\,,
  \label{eq:fourckmamps}
\eeq
where we have separated the $CKM$ factors $\lambda_{c,u}^{(*)}$ and called the rest 
$F_{c,u}$
which include strong phases as well as decay constants and form
factors (if factorization is assumed). 
We assume that the
$CP$ violation is solely through the weak phases that appear in (\ref{eq:fourckmamps});
as a consequence we can show that (see Appendix)
\beq
      F_c = \bar F_c\,,\quad F_u = \bar F_u\,.
  \label{eq:dpicpfa}
\eeq
We then have
\beq
   |a| = |\bar a|\,,\quad |b| = |\bar b|\,.
\eeq

Using (\ref{eq:povq}) and (\ref{eq:fouramps}) as well as (\ref{eq:dpicpfa}), 
the value of $\bar\rho$ defined in (\ref{eq:rhodef}) is then
\beq
   \bar\rho \equiv {p\, b\over q\,\bar a} 
    = {p\, \lambda_u^* F_u\over q\,  \lambda_c F_c} =
   -{V_{td}^* V_{tb} \over V_{td} V_{tb}^*}
   {V_{ub}^*V_{cd}\over V_{cb}V_{ud}^*}
   {F_u\over F_c} \equiv r e^{i\phi_{\bar\rho}}\,.
\eeq
where we have defined $r\equiv |\bar\rho|$ and $\phi_{\bar\rho} \equiv \arg \bar\rho$.
With the definitions of $\phi_1$ and $\phi_3$
\beq
   \phi_1 \equiv \arg{V_{cd}V_{cb}^*\over - V_{td}V_{tb}^*}\,,\quad
   \phi_3 \equiv \arg{V_{ud}V_{ub}^*\over - V_{cd}V_{cb}^*}\,,
  \label{eq:phi13def}
\eeq
we have
\beq
  \arg\left( -{V_{td}^* V_{tb} \over V_{td} V_{tb}^*}
             {V_{ub}^*V_{cd}\over V_{cb}V_{ud}^*} \right) =
                 \arg\left(
                       {V_{cd}V_{cb}^*\over - V_{td}V_{tb}^*}
                        {V_{cd}V_{cb}^*\over - V_{td}V_{tb}^*}
                        {V_{ud}V_{ub}^*\over - V_{cd}V_{cb}^*}\right)
     = 2\phi_1 + \phi_3\,.
\eeq
Then, $\bar\rho$ can be written as
\beq
   \bar\rho =  r e^{i(\phi_w +\delta)}\,.
  \label{eq:rhoval}
\eeq
with
\beq
    \phi_w \equiv 2\phi_1+\phi_3\,,\quad  \delta \equiv \arg {F_u\over F_c}\,.
\eeq
Similarly, one obtains
\beq
       \rho =  r e^{-i(\phi_w - \delta)}\,.
  \label{eq:rhobarval}
\eeq
Note that we have $|\rho| = |\bar\rho| =r$. The value of $r$ is roughly
\beq
    r = \left|{V_{ub}^*V_{cd}\over V_{cb}V_{ud}^*}{F_u\over F_c}\right|
  \sim  0.4 \lambda^2 \sim 0.02\,. \quad (\lambda\sim0.22:\hbox{Cabibbo factor})\,.
\eeq

With
$\Im\rho = - r\sin(\phi_w-\delta)$ and 
$\Im\bar\rho =  r\sin(\phi_w+\delta)$, the four decay rates
(\ref{eq:fourrates}) becomes
\beq
 \begin{array}{rc@{\tbst}l}
    \Gamma_{B^0\to D^-\pi^+}(t) &=& |a|^2
      {e^{-\gamma t}\over2} \left[ (1+r^2) + (1-r^2)\cos\delta mt 
                         - 2r\sin(\phi_w-\delta)\,\sin\delta mt \right] \\
    \Gamma_{\bar B^0\to D^+\pi^-}(t) &=& |a|^2
      {e^{-\gamma t}\over2} \left[ (1+r^2) + (1-r^2)\cos\delta mt 
                         +  2r\sin(\phi_w+\delta) \,\sin\delta mt \right] \\
    \Gamma_{B^0\to D^+\pi^-}(t) &=& |a|^2
      {e^{-\gamma t}\over2} \left[ (1+r^2) - (1-r^2)\cos\delta mt 
                         -  2r\sin(\phi_w+\delta) \,\sin\delta mt \right] \\
    \Gamma_{\bar B^0\to D^-\pi^+}(t) &=& |a|^2
      {e^{-\gamma t}\over2} \left[ (1+r^2) - (1-r^2)\cos\delta mt 
                         +  2r\sin(\phi_w-\delta)\,\sin\delta mt \right]
 \end{array}\,.
  \label{eq:ratesdpi}
\eeq
where have used $|a| = |\bar a|$.
Note that $\Gamma_{\bar B^0\to D^-\pi^+}(t)$ (suppressed) is obtained from 
$\Gamma_{B^0\to D^-\pi^+}(t)$ (favored) and $\Gamma_{B^0\to D^+\pi^-}(t)$
(suppressed) is obtained from $\Gamma_{\bar B^0\to D^+\pi^-}(t)$ (favored) by the
transformation (\ref{eq:supfavtransf}), and within the two suppressed modes
and within the favored modes, the expresssions are related by
(\ref{eq:BBbartransf}) namely $\phi_w \leftrightarrow -\phi_w$.

The $CP$ violating parameters that can be
extracted from these distributions are
\beq
       r\sin(\phi_w-\delta)\quad{\rm and}\quad
       r\sin(\phi_w+\delta)\,.
\eeq
Note that the two extractable
paramters are always multiplied with $r$, and the value of $r$ cannot
be obtained by the fit.
As discussed earlier, the corresponding distributions on $\Upsilon(4S)$ are obtained by
replacements $\gamma t \to \gamma |\Delta t|$ and $\delta m t \to \delta m \Delta t$.
The first paramter $r\sin(\phi_w-\delta)$ can be obtained through
the asymmetry between positive and negative $\Delta t$ of
$\Gamma_{\ell^-,D^-\pi^+}(\Delta t)$ (favored)
or $\Gamma_{\ell^+,D^-\pi^+}(\Delta t)$ (suppressed), and the second paramter
$r\sin(\phi_w+\delta)$ is similarly obtained through
$\Gamma_{\ell^+, D^+\pi^-}(t)$ (favored) or
$\Gamma_{\ell^-, D^+\pi^-}(t)$ (suppressed). This feature that single
mode can give a $CP$ violating parameter 
through asymmetry between positive and negative $\Delta t$
is unique to $\Upsilon(4S)$. In fact, most of the information on $CP$ violation
is in such asymmetries. If we define
\beq
    \delta\Gamma_X(|\Delta t|) \equiv
   \Gamma_X(\Delta t) - \Gamma_X(-\Delta t)\,,
\eeq
we have
\beqa
   -\delta\Gamma_{\ell^-,D^-\pi^+}(|\Delta t|) = 
   \delta\Gamma_{\ell^+,D^-\pi^+}(|\Delta t|) &=&
    N r\sin(\phi_w-\delta) e^{-\gamma t} \sin(\delta m t) \,,\\
   \delta\Gamma_{\ell^+,D^+\pi^-}(|\Delta t|) = 
   -\delta\Gamma_{\ell^-,D^+\pi^-}(|\Delta t|) &=&
    N r\sin(\phi_w+\delta) e^{-\gamma t} \sin(\delta m t) \,,
\eeqa
where $N$ is a common normalization factor which is known.

Now we derive the corresponding time-integrated expressions. 
We use following integrals. 
\begin{eqnarray}
&\,\int_{0}^{\infty}e^{-\gamma t}\, dt&  = \frac{1}{\gamma} \,, \\
&\,\int_{0}^{\infty}e^{-\gamma t}\sin \delta mt\, dt& 
                     = \frac{1}{\gamma}\frac{x}{1+x^{2}} \,, \\
&\,\int_{0}^{\infty}e^{-\gamma t}\cos \delta mt\, dt &=
\frac{1}{\gamma}\frac{1}{1+x^{2} } \,.
\label{eq:integral}
\end{eqnarray}
Here $x={\delta m}/{\gamma}$. 
The time-integrated decay rates become 
\begin{eqnarray}
\Gamma(B^{0} \rightarrow D^{-} \pi^+) &=& 
\frac{|a|^2}{2\gamma}[(1+r^2)+\frac{1-r^2}{1+x^2}-
\frac{2rx}{1+x^2}\sin(\phi_w -\delta)] \nonumber \\  
\Gamma(\bar B^{0} \rightarrow D^{+} \pi^-) &=& 
 \frac{|a|^{2}}{2\gamma}
 [(1+r^2)+\frac{1-r^2}{1+x^2}+ \frac{2rx}{1+x^2}\sin(\phi_w +\delta)]
    \nonumber\\
\Gamma(B^{0} \rightarrow D^{+} \pi^-) &=& 
 \frac{|a|^{2}}{2\gamma}
[(1+r^2)-\frac{1-r^2}{1+x^2}-\frac{2rx}{1+x^2}\sin(\phi_w +\delta) ] 
\nonumber \\
\Gamma(\bar B^{0} \rightarrow D^{-} \pi^+) &=& 
\frac{|a|^2}{2\gamma}[(1+r^2)-\frac{1-r^2}{1+x^2}+
\frac{2rx}{1+x^2}\sin(\phi_w-\delta)] 
  \label{eq:rintdpi}
\end{eqnarray}
If we set $\delta = 0$ for simplicity, we see that the information on
$\sin(\phi_w)$ is in the asymmetry between the top two
rates (the favored modes) or in the asymmetry between the bottom
two rates (the suppressed modes). The absolute amount of
the differnce is the same for both cases, but the total rate is about
5 times larger for the favored modes compared to the suppressed modes.
It means that the significance (number of sigmas) is $\sqrt5$ times
smaller for the favored modes.
Thus, most of the information is contained in the suppressed modes.

The expressions (\ref{eq:ratesdpi}) and (\ref{eq:rintdpi}) are valid also for
$f = D^{*-}\pi^+$, $D^{-}\rho^+$. When there are
more than one polarization states as in $D^{*-}\rho^+$, there is extra
effect due to interferences between different polarization states, which
we will discuss next.

\subsection{$B^0  \to D^{*-} \rho^+$}

This mode was first stdied in detail in Ref.~\cite{LSS}.
We first note that the expressions for the time dependent amplitudes
(\ref{eq:timeamp4}) are still valid when applied to each polarization
state:
\beq
  \begin{array}{rc@{\tbst}lrcl}
    A_{B^0\to f_\lambda}(t) &=&    
      e^{-{\gamma\over2}t}\, \left( a_\lambda 
       \cos{\delta m\, t\over2} - {q\over p}\bar b_\lambda i\sin{\delta m\, t\over2}
     \right) &=& 
      e^{-{\gamma\over2}t}\, a_\lambda \left(
     \cos{\delta m\, t\over2} - \rho_\lambda\, i\sin{\delta m\, t\over2}
     \right)   \\
    A_{\bar B^0\to \bar f_\lambda}(t) &=&    
      e^{-{\gamma\over2}t}\, \left( \bar a_\lambda 
       \cos{\delta m\, t\over2} - {p\over q} b_\lambda i\sin{\delta m\, t\over2}
     \right) &=& 
      e^{-{\gamma\over2}t}\, \bar a_\lambda \left(
     \cos{\delta m\, t\over2} - \bar \rho_\lambda\, i\sin{\delta m\, t\over2}
     \right) \\
    A_{B^0\to \bar f_\lambda}(t)&=&    
      e^{-{\gamma\over2}t}\, \left( b_\lambda 
       \cos{\delta m\, t\over2} - {q\over p}\bar a_\lambda i\sin{\delta m\, t\over2}
     \right) &=& 
      e^{-{\gamma\over2}t}\, \bar a_\lambda\left(
      \bar \rho_\lambda \cos{\delta m\, t\over2} - i\sin{\delta m\, t\over2}
     \right)  \\
    A_{\bar B^0\to f_\lambda}(t) &=&    
      e^{-{\gamma\over2}t}\, \left( \bar b_\lambda 
       \cos{\delta m\, t\over2} - {p\over q} a_\lambda i\sin{\delta m\, t\over2}
     \right)&=& 
      e^{-{\gamma\over2}t}\, a_\lambda \left(
     \rho_\lambda \cos{\delta m\, t\over2} - i\sin{\delta m\, t\over2}
     \right) 
  \end{array}
  \label{eq:timeampvv}
\eeq
where
\beq
  \begin{array}{r@{\tbst}cl}
          a_\lambda &\equiv& Amp(B^0\to f_\lambda) = \lambda_c^* F_{c\lambda}\\
     \bar a_\lambda &\equiv& Amp(\bar B^0\to \bar f_\lambda)
           = \lambda_c \bar F_{c\lambda} \\
          b_\lambda &\equiv& Amp(B^0\to \bar f_\lambda) = \lambda_u^* F_{u\lambda} \\
     \bar b_\lambda &\equiv& Amp(\bar B^0\to f_\lambda) = \lambda_u \bar F_{u\lambda} 
  \end{array}\,,
  \label{eq:fouramppol}
\eeq
and
\beq
  \rho_\lambda\equiv {q\, \bar b_\lambda\over p\, a_\lambda}\,,\quad
  \bar\rho_\lambda\equiv {p\, b_\lambda\over q\, \bar a_\lambda}\,.
  \label{eq:rhodeflam}
\eeq

Each of (\ref{eq:timeampvv})
gives the polarization amplitudes to a given final state at time $t$.
Then, the angular distribution of pure $B^0$ at $t=0$
decaying to $f$ at time $t$ is simply obtained by
replacing $H_\lambda$ or $A_\lambda$ by $A_{B^0\to f_\lambda}(t)$ in 
(\ref{eq:angdisheldrho}) or (\ref{eq:angtrdstrho}). For $B^0(t=0)\to f$, for
example, the time-dependent angular distribution is given by (\ref{eq:angtrdstrho})
with the replacement
\beqa
   |A_\lambda|^2 &\to& |A_{B^0\to f_\lambda}(t)|^2\,, \nonumber \\
   \Re(A_\|^* A_0) &\to& \Re(A^*_{B^0\to f_\|}(t)A_{B^0\to f_0}(t))\,, \nonumber\\
   \Im(A_0^* A_\bot) &\to& \Im(A^*_{B^0\to f_0}(t)A_{B^0\to f_\bot}(t))\,,\nonumber \\
   \Im(A_\|^* A_\bot) &\to& \Im(A^*_{B^0\to f_\|}(t)A_{B^0\to f_\bot}(t)) \,.
  \label{eq:coefrepl}
\eeqa
Or the decay amplitudes are obtained from (\ref{eq:dpiampt})
by the same replacement:
\beqa 
  A_{B^0\to f}(\Omega,t) &=&
     \sum_\lambda  e^{-{\gamma\over2}t}\, a_\lambda \left(
     \cos{\delta m\, t\over2} - \rho_\lambda\, i\sin{\delta m\, t\over2}\right)
     g_\lambda (\Omega) \nonumber \\
  A_{\bar B^0\to \bar f}(\Omega,t) &=&
     \sum_\lambda  e^{-{\gamma\over2}t}\, \bar a_\lambda \left(
     \cos{\delta m\, t\over2} - \bar\rho_\lambda\, i\sin{\delta m\, t\over2}\right)
     g_\lambda (\Omega) \nonumber \\
  A_{B^0\to \bar f}(\Omega,t) &=&
     \sum_\lambda  e^{-{\gamma\over2}t}\, \bar a_\lambda \left(
      \bar\rho_\lambda\, \cos{\delta m\, t\over2} -i\sin{\delta m\, t\over2}\right)
     g_\lambda (\Omega) \nonumber \\
  A_{\bar B^0\to f}(\Omega,t) &=&
     \sum_\lambda  e^{-{\gamma\over2}t}\, a_\lambda \left(
     \rho_\lambda\,\cos{\delta m\, t\over2} -  i\sin{\delta m\, t\over2}\right)
     g_\lambda (\Omega)\,,
 \label{eq:dstrhoamps}
\eeqa
where $\Omega\equiv (\phi_{tr},\theta_{tr},\psi)$ or $(\chi,\theta,\psi)$,
and $g_\lambda$'s are given
by (\ref{eq:gmtrdstrho}) or (\ref{eq:gmhldstrho}). 
Here, the final states are $f \equiv D^{*-}\rho^+$ and $\bar f \equiv D^{*+}\rho^-$.
Note that one could use
angles $(\phi_{tr},\theta_{tr},\psi)$ or $(\chi,\theta,\psi)$ for the tranversity
amplitudes (for that matter, for the helicity amplitudes also - we just
have not provided $g(\phi_{tr},\theta_{tr},\psi)$ for the helicity amplitudes).

Since $F_{u\lambda}$ and $F_{c\lambda}$ are nothing but the polarization
amplitudes apart from the $CP$ violating phases, they themselves should
satisfy the $CP$ relations (\ref{eq:cphel}) and (\ref{eq:cptrans}) 
(see Appendix):
\beq
    \bar F_{q\lambda} = F_{q-\lambda}\quad(\hbox{helicity})\,,
  \label{eq:cphelf}
\eeq
\beq
     \bar F_{q\|} = F_{q\|}\,,\quad \bar F_{q0} = F_{q0}\,,\quad
     \bar F_{q\bot} = -F_{u\bot}\,,\quad(\hbox{tranversity})
  \label{eq:cptransf}
\eeq
where $q=u$ or $c$.

\subsubsection{Helicity basis}

With the relations (\ref{eq:cphelf})
for helicity basis, the decay amplitudes can be written as
\beq
  \begin{array}{r@{\tbst}clrcl}
          a_\lambda &=& \lambda_c^* F_{c\lambda}\,,\quad
     \bar a_{-\lambda} &=& \lambda_c F_{c\lambda}\,, \\
          b_\lambda &=& \lambda_u^* F_{u\lambda} \,,\quad
     \bar b_{-\lambda} &=& \lambda_u F_{u\lambda} \,.
  \end{array}\quad (\hbox{helicity})
\eeq
This gives
\beq
    |a_\lambda| = |\bar a_{-\lambda}|\,,\quad
    |b_\lambda| = |\bar b_{-\lambda}|\,.
\eeq
And the parameters $\rho_\lambda$ and $\bar\rho_\lambda$ becomes
\beq
   \rho_\lambda = {q\over p}{\lambda_u   F_{u -\lambda}
                        \over\lambda_c^* F_{c  \lambda}}\,,\quad
   \bar\rho_{-\lambda} = {p\over q}{\lambda_u^* F_{u -\lambda}
                            \over\lambda_c   F_{c \lambda}}\,.
\eeq
Then, the same procedure that led to (\ref{eq:rhoval}) and (\ref{eq:rhobarval})
allows one to write
\beq
   \rho_\lambda = r_\lambda e^{i(\phi_w+\delta_\lambda)}\,,\quad
   \bar\rho_{-\lambda} = r_\lambda e^{-i(\phi_w -\delta_\lambda)}\,,\quad
\eeq
where
\beq
     |\rho_\lambda| = |\bar\rho_{-\lambda}| \equiv r_\lambda\,,
\eeq
and
\beq
     \delta_\lambda \equiv \arg{F_{u-\lambda} \over F_{c\lambda}}\,.
\eeq

\subsubsection{Transversity basis}

Using the relations (\ref{eq:cptransf}) for
transversity, we can write
\beq
  \begin{array}{r@{\tbst}clrcl}
          a_\lambda &=& \lambda_c^* F_{c\lambda}\,,\quad
     \bar a_\lambda &=& \xi_\lambda \lambda_c F_{c\lambda}\,, \\
          b_\lambda &=& \lambda_u^* F_{u\lambda} \,,\quad
     \bar b_\lambda &=& \xi_\lambda \lambda_u F_{u\lambda} \,,
  \end{array}\quad (\hbox{transversity})
\eeq
where
\beq
     \xi_\lambda = \left\{\begin{array}{rl}
                           1 &\quad (\lambda = \para, 0) \\
                          -1 & \quad(\lambda = \bot) 
                          \end{array}\right. \,.
  \label{eq:xidef}
\eeq
We will use the tranveristy basis for the rest of this section.
Clearly, we have
\beq
    |a_\lambda| = |\bar a_\lambda|\quad{\rm and}\quad
    |b_\lambda| = |\bar b_\lambda|\,,
\eeq
and the procedure semilar to that led to (\ref{eq:rhoval}) and (\ref{eq:rhobarval})
gives
\beq
   \rho_\lambda = \xi_\lambda r_\lambda e^{-i(\phi_w - \delta_\lambda)}
\,,\quad
  \bar \rho_\lambda = \xi_\lambda r_\lambda 
    e^{i(\phi_w + \delta_\lambda)}\,,
\eeq
where
\beq
   |\rho_\lambda| = |\bar\rho_\lambda| \equiv r_\lambda\,,
\eeq
and
\beq
    \delta_\lambda \equiv \arg{F_{u\lambda}\over F_{c\lambda}}\,.
\eeq

Let's evaluate the expilict decay rates; namely, the coefficients
given in (\ref{eq:coefrepl}).
Note that $\rho_\lambda$ and $\bar\rho_\lambda$ are related by
$\phi_w \leftrightarrow -\phi_w$. Togethter with (\ref{eq:supfavtransf}),
all we need is to evaluate one of the four modes which we take to be
the favored mode $B^0\to D^{*-}\rho^+$. 
Calculation is straightforward and we obtain (apart from the
common factor $e^{-\gamma t}/2$)
\beqa
 |A_\lambda|^2 &\to&
    |a_\lambda|^2 \Big[
    (1 + r_\lambda^2) + (1 - r_\lambda^2) \cos\delta mt \nonumber\\
   && \hspace{1in} - 2 \xi_\lambda r_\lambda 
     \sin(\phi_w - \delta_\lambda)\sin\delta mt \Big] \,, \nonumber \\
  \Re(A_\|^* A_0) &\to&
    \Big[ \Re(a_\|^* a_0) (1 + r_\| r_0\cos(\delta_\|-\delta_0)) 
         +\Im(a_\|^* a_0) r_\| r_0 \sin(\delta_\| - \delta_0) \Big] \nonumber\\
    &+& \Big[\Re(a_\|^* a_0)  (1 - r_\| r_0\cos(\delta_\|-\delta_0))
       - \Im(a_\|^* a_0) r_\| r_0 \sin(\delta_\| - \delta_0)\Big]
              \cos\delta mt  \nonumber \\
   &-& \Big[\Re(a_\|^* a_0) (r_\| \sin(\phi_w-\delta_\|) + r_0 
              \sin(\phi_w-\delta_0)\big)
         \nonumber\\
  &&\; +  \Im(a_\|^* a_0) 
    (r_\| \cos(\phi_w-\delta_\|) - r_0 \cos(\phi_w-\delta_0))\Big]\sin\delta mt\,, \\
  \Im(A_e^* A_\bot) &\to&
    \Big[\Im(a_e^* a_\bot)(1 - r_e r_\bot \cos(\delta_e - \delta_\bot)) 
       + \Re(a_e^* a_\bot)r_e r_\bot \sin(\delta_e - \delta_\bot) \Big] \nonumber\\
  &+& \Big[\Im(a_e^* a_\bot)(1 + r_e r_\bot \cos(\delta_e - \delta_\bot))
       - \Re(a_e^* a_\bot)r_e r_\bot \sin(\delta_e - \delta_\bot) \Big] 
         \cos\delta mt  \nonumber \\
   &-& \Big[ \Im(a_e^* a_\bot)  (r_e \sin(\phi_w-\delta_e) - r_\bot 
              \sin(\phi_w-\delta_\bot)\big)
         \nonumber\\
  &&\; - \Re(a_e^* a_\bot) (
    r_e \cos(\phi_w-\delta_e) + 
     r_\bot \cos(\phi_w-\delta_\bot))\Big]\sin\delta mt\,, \nonumber 
\eeqa
where $\lambda = (\|, 0, \bot)$, $e = (\|, 0)$, and the suppressed modes for the
same final states are obtained by the transformation $\delta mt \to
\delta mt + \pi$ or (\ref{eq:supfavtransf}), and among the two suppressed
or among the two favored modes, the $B^0$ decay and the $\bar B^0$ decay
are related by $\phi_w \leftrightarrow -\phi_w$. The distribution
(\ref{eq:angtrdstrho}) with these replacements then gives the desired
time-dependent angular distributions.

\subsubsection{Fit parameters}

Squares of the amplitudes (\ref{eq:dstrhoamps}) give the rates, and
with complex functions in programing language, these expressions are all needed
to perform the fit. The fit parameters are $a_\lambda$, $\bar a_\lambda$, 
$\rho_\lambda$, and $\bar\rho_\lambda$. Note that only the relative phases matter
among $a_\lambda$ and among $\bar a_\lambda$; namely, one can set $a_0$ = real and
$\bar a_0$ = real, for example. In addition, 
$|a_\lambda| = |\bar a_{-\lambda}|$ (helicity) or 
$|a_\lambda| = |\bar a_\lambda|$ (transversity)
reduces the number of degrees of freedom by 3 in each basis.
Furthermore, there are phase relations in
\beq
   {\bar a_{-\lambda}\over\bar a_0} = {a_{\lambda}\over a_0}\quad
   (\lambda=\pm1)\,,\quad{\rm or}\quad
   {\bar a_\lambda\over\bar a_0} = \xi_\lambda{a_\lambda\over a_0}\quad
   (\lambda=\|,\bot)\,,
\eeq
which reduces 2 degrees of freedom.
Thus, there are 5 degrees of freedom in $a_\lambda$ and $\bar a_\lambda$
including the overall normalizations. One may parametrize, for example, as
\beq
  \begin{array}{lcl}
   a_{+1} &=& |a_{+1}| e^{i\phi_{+1}} \\
   a_0  &=& |a_0|  \\
   a_{-1} &=& |a_{-1}| e^{i\phi_{-1}}
  \end{array}\,,\quad
  \begin{array}{lcl}
   \bar a_{+1} &=& |a_{-1}| e^{i\phi_{-1}} \\
   \bar a_0    &=& |a_0|  \\
   \bar a_{-1} &=& |a_{+1}| e^{i\phi_{+1}}
  \end{array}\quad\hbox{(helicity)}\,,
\eeq
or,
\beq
  \begin{array}{lcl}
   a_\| &=& |a_\|| e^{i\phi_\|} \\
   a_0  &=& |a_0|  \\
   a_\bot &=& |a_\bot| e^{i\phi_\bot}
  \end{array}\,,\quad
  \begin{array}{lcl}
   \bar a_\|   &=& |a_\|| e^{i\phi_\|} \\
   \bar a_0    &=& |a_0|  \\
   \bar a_\bot &=& -|a_\bot| e^{i\phi_\bot}
  \end{array}\quad\hbox{(transversity)}\,.
\eeq
Also $\rho_\lambda$ and $\bar\rho_\lambda$ are constrained
by the expression (\ref{eq:rhodeflam}); namely, we actually fit
$r_\lambda$, $\delta_\lambda$, and $\phi_w$, 
which amounts to 7 degrees of freedom. The total number of degrees freedom is
thus $5+7=12$ including the overall normalization.

\subsection{Time dependent angular distribution for $\Psi K^{*0}$}

The only one relevant final state to be considered is $\Psi K^{*0}$ 
where $K^{*0}$ decays to $K_S \pi^0$. 
We denote the final state as $f_\lambda = (\Psi K_S^{*0})_\lambda$,
where $\lambda$ could be for helicity basis or transveristy basis.
The particle assignments are
\beq
   a = \Psi\,,\quad 
   a_1 = \ell^+\,,\quad a_2 = \ell^-\,,\quad
   b = K^{*0}\,,\quad
   b_1 = K_S\,,\quad a_2 = \pi^0\,.
\eeq
All we need is the amplitudes for each polarization
(helicty basis or transversity basis) at time $t$ when 
the $B$ meson was pure $B^0$ or $\bar B^0$ at $t=0$.
Then, we can use the distributions (\ref{eq:angdishel}) and
(\ref{eq:angdistr}) to obtain the angular distribution at that time.
Incoherent sum over the two possible helicity states of the $\Psi$
decay is already taken into account in those angular distributions.

Since we are dealing with only one final state (apart from polarization), 
the first and the
last of (\ref{eq:timeamp4}) will do:
\beq
  \begin{array}{rc@{\tbst}l}
    A_{B^0\to f_\lambda}(t) &=& 
      e^{-{\gamma\over2}t} a_\lambda \left(
     \cos{\delta m\, t\over2} - \rho_\lambda\, i\sin{\delta m\, t\over2}
     \right) \\
    A_{\bar B^0\to f_\lambda}(t) &=& 
      e^{-{\gamma\over2}t} a_\lambda \left(
     \rho_\lambda\cos{\delta m\, t\over2} - i\sin{\delta m\, t\over2}
     \right)
  \end{array}\,,
\eeq
where
\beq
     a_\lambda = Amp(B^0\to (\Psi K^{*0}_S)_\lambda)\,,\quad
     \bar b_\lambda = Amp(\bar B^0\to (\Psi K^{*0}_S)_\lambda)\,,\quad
    \rho_\lambda = {q\, \bar b_\lambda\over p\,a_\lambda}\,.
\eeq
The $\rho_\lambda$ parameter is then
\beq
   \rho_\lambda = {q\over p}
 {\bra K_S|\bar K^0\ket \over \bra K_S| K^0\ket}
 {\bra (\Psi \bar K^{*0})_\lambda| H_{\rm eff} | \bar B^0\ket \over
  \bra (\Psi K^{*0})_\lambda| H_{\rm eff} | B^0\ket }\,.
\eeq
Eq. (\ref{eq:povq}) gives $q/p$, and using (\ref{eq:povqK}) of 
Appendix,
\beq
   {\bra K_S|\bar K^0\ket \over \bra K_S| K^0\ket} = {-q_K^*\over p_K^*}
   = {V_{cs}V_{cd}^*\over V_{cs}^*V_{cd}}
\eeq
can be obtained as in the case of $B$ where we have ignored the small
deviation of $|q_K/p_K|$ from unity.
Assuming that the
color-suppressed tree diagram dominates the amplitudes $a$ and $\bar b$,
or assumming that penguin and other contributions do not modify the
weak phase significantly,
\beq
   \bra (\Psi K^{*0})_\lambda| H_{\rm eff} | B^0\ket
      = V_{cb}^*V_{cs} F_\lambda\,,\quad 
   \bra (\Psi \bar K^{*0})_\lambda| H_{\rm eff} | \bar B^0\ket
      = V_{cb}V_{cs}^* \bar F_\lambda\,.
   \label{eq:PsiKstamps}
\eeq
By the similar argument that led to 
the $CP$ relations (\ref{eq:cphelf}) and
(\ref{eq:cptransf}), $F_\lambda$ and $\bar F_\lambda$ are related by
\beq
    \bar F_{\lambda} = F_{-\lambda}\quad(\hbox{helicity})\,,
\eeq
\beq
     \bar F_{\|} = F_{\|}\,,\quad \bar F_{0} = F_{0}\,,\quad
     \bar F_{\bot} = -F_{\bot}\,,\quad(\hbox{tranversity})
\eeq

Let's use the transversity basis for the rest of this section.
Then, the amplitudes given by (\ref{eq:PsiKstamps}) togehter
with the $CP$ relation above gives
\beq
 {\bra (\Psi \bar K^{*0})_\lambda| H_{\rm eff} | \bar B^0\ket \over
  \bra (\Psi K^{*0})_\lambda| H_{\rm eff} | B^0\ket }
  = \xi_\lambda {V_{cb}V_{cs}^*\over V_{cb}^*V_{cs}}
   \quad\hbox{(transversity)}\,,
\eeq
where $\xi_\lambda$ is the sign defined by (\ref{eq:xidef}).
Combining all ingredients, $\rho_\lambda$ becomes
\beq
   \rho_\lambda = \left(-{V_{tb}^* V_{td}\over V_{tb} V_{td}^*}\right)
   \left({V_{cs}V_{cd}^*\over V_{cs}^*V_{cd}}\right)
   \left(\xi_\lambda {V_{cb}V_{cs}^*\over V_{cb}^*V_{cs}}\right) =
   -\xi_\lambda\left( {V_{cd}V_{cb}^*\over -V_{td}V_{tb}^*}\right)^*\Bigg/
    \left( {V_{cd}V_{cb}^*\over -V_{td}V_{tb}^*}\right)\,.
\eeq
With the definition of $\phi_1$ (\ref{eq:phi13def}), we can write
\beq
   \rho_\lambda = -\xi_\lambda e^{-2i\phi_1}
   \quad\hbox{(transversity)}\,.
\eeq
Recall that the value of $\rho$ for the gold-plated $\Psi K_S$ final
state was $e^{-2i\phi_1}$; namely, the transverse polarization $A_\bot$
has the same time-dependent $CP$ asymmetry as the $\Psi K_S$
final state, and $A_\|$ and $A_0$ states have the $CP$ asymmetry
opposite to that of $\Psi K_S$. These arguments are valid when a given
polarization state dominates the final state and when integrated over
the angular distribution. 

The angular distribution is given by the expression (\ref{eq:angdistr})
with the coefficient replaced according to (\ref{eq:coefrepl}).
Explicitly,
\beqa
  |A_\lambda|^2 &\to& 
     |a_\lambda|^2 (1 \pm \xi_\lambda\sin2\phi_1\sin\delta mt) \nonumber\\
  \Re(A_\|^* A_0) &\to&
    \Re(a_\|^* a_0) (1 \pm \sin2\phi_1 \sin\delta mt) \\
  \Im(A_e^* A_\bot) &\to&
    \pm \Im(a_e^* a_\bot)\cos\delta mt \mp \Re(a_e^*a_\|)\cos2\phi_1\sin\delta mt
     \nonumber
\eeqa
where the upper sign is for $B^0\to\Psi K^{*0}_S$,
the bottom sign is for $\bar B^0\to\Psi K^{*0}_S$, 
$\lambda = (\|,0,\bot)$ and $e$ stands for $\|$ or $0$.
They are related by the transformation (\ref{eq:supfavtransf}) as
expected. 
Note that $\cos2\phi_1$ can be obtained by these angular distributions, which
helps to resolve the discrete ambiguity of $\phi_1$.

\section{Appendix}

\subsection{$CP$ relations}

We will hereby derive the $CP$ relations (\ref{eq:cphel})
and (\ref{eq:cphelf}). Suppose the effective Hamiltonian 
commutes with $CP$:
\beq
   I H_{\rm eff} I^\dag = H_{\rm eff}\,,
   \label{eq:cpheff}
\eeq
where
\beq
    I\,\equiv CP\,.
\eeq
For example, the Hamiltonian for the tree diagram of $B^0\to D^-\pi^+$ is
(up to a constant)~\footnote{What we are calling $H_{\rm eff}$ is 
actually the $S$ operator.}%
\beq
      H_{\rm eff} = \lambda_c h_c + (h.c.)
 \label{eq:heffform}
\eeq
with
\beq
      h_c \equiv \int_{-T}^T dt \int d^3x (\bar c b)_\mu(\bar d u)^\mu \,,
\eeq
where $ (\bar q q')_\mu$ is a color-singlet $V-A$ current which is a function of
space-time:
\beq
     (\bar q q'(x))_\mu \equiv \bar q^a(x) \gamma_\mu (1-\gamma_5) q^{\prime a}(x)\,.
  \qquad (x = (t,\vec x))
\eeq
with $a$ being the color index. 
The $CKM$ factor $\lambda_c$ is defined in
(\ref{eq:fouramps}). The $CP$ phases of quark fields are taken as
\beq
     \eta_q = 1\,,
\eeq
where the $CP$ phase is defined by
\beq
    I \,|q_{\vec p\sigma}\ket = \eta_q |\bar q_{-\vec p, \sigma}\ket\,,
\eeq
where $\vec p$ is the momentum and $\sigma$ is the spin component along $z$.
With this choice of $CP$ phase, one can show that (see, for example, Ref~\cite{TDLee})
\beq
    I (\bar q q'(x))_\mu I^\dag = -{(\bar q q'(x'))^\mu}^\dag = -(\bar q' q(x'))^\mu 
   \qquad (x' = (t,-\vec x))\,.
\eeq
Note that the Lorentz index $\mu$ changed from subscript to superscript.
We then have after the integration over space
\beq
   I\,\int d^3x (\bar c b)_\mu(\bar d u)^\mu\, I^\dag 
     = \left[ \int d^3x (\bar c b)^\mu(\bar d u)_\mu\,\right]^\dag\,,
\eeq
which leads to
\beq
   I\, h_c \,I^\dag = h_c^\dag\,.
 \label{eq:hcpparta}
\eeq
 Similarly, we can show
\beq
    I\,h_c^\dag I^\dag = h_c\,.
 \label{eq:hcppartb}
\eeq
This makes (\ref{eq:cpheff}) hold if $\lambda_c$ is real.
In general, $H_{\rm eff}$ includes strong interaction that results in
phase shifts. Still, it can be writen in the form (\ref{eq:heffform})
and that it would be invariant under $CP$ if the $CKM$ factors are
real; namely, (\ref{eq:hcpparta}) and (\ref{eq:hcppartb}) are
satisfied.

The {\it helicity} states of $B\to a+b$ transforms under $CP$ as
(see, for example, Ref~\cite{JW})
\beq
   I \,|JM,\lambda_a\lambda_b;ab\ket = 
   \eta_a \eta_b (-)^{J-s_a-s_b}
   |JM,-\lambda_a-\lambda_b;\bar a\bar b\ket\,,
  \label{eq:CP2bgen}
\eeq
where $J=0$ and $s_{a,b}=1$ for our case, and
$\eta_a$ and $\eta_b$ are the $CP$ phases of $a$ and $b$ respectively:
\beq
    CP |a;\vec p,\sigma\ket = \eta_a |\bar a;-\vec p,\sigma\ket\,,\quad
    CP |b;\vec p,\sigma\ket = \eta_b |\bar b;-\vec p,\sigma\ket\,,
\eeq
where $\sigma$ is the $z$-component of spin.
If $a$ or $b$ are not self-conjugate, then their $CP$ phases are
cancelled when the value of $\rho$ is calculated or relation between
$\rho_\lambda$ and $\bar\rho_\lambda$ is evaluated. 
For a self-conjuagte particle, the
$CP$ phase does matter. However,
$CP$ of relvant spin-1 particles, such as
any known spin-1 $(c\bar c)$ states, $\rho^0$, $a_1^0$, $\omega$,
$\phi$, etc. are all $+1$. Thus, we take $\eta_a\eta_b$ to be $+1$
keeping in mind that if any of the spin-1 particles are self-conjugate
and $CP-$ then it has to be included in the sign.
Thus, in terms of our short notation, the above relation becomes
\beq
     I\, |f_\lambda\ket =|\bar f_{-\lambda}\ket\,.
  \label{eq:cpflam}
\eeq

Then, the helicity amplitude transforms as (with $\eta_B=1$)
\beqa
    H_\lambda &=& \bra f_\lambda | H_{\rm eff} | B^0\ket \\
    &=& \underbrace{\bra f_\lambda |\,I^\dag}_{\dspl \bra \bar f_{-\lambda}|}\,
    \underbrace{I\, H_{\rm eff}\,I^\dag}_{\dspl H_{\rm eff}} \,
    \underbrace{I\,| B^0\ket }_{\dspl \eta_B|\bar B^0\ket} = \bar H_{-\lambda}\,,
\eeqa
which proves  (\ref{eq:cphel}).

The proof of the relation (\ref{eq:cphelf}) starts from realizing that
the relevant effective Hamiltonian can be written as
\beq
      H_{\rm eff} = (\lambda_c h_c + \lambda_u h_u) + 
       (\lambda_c^* h_c^\dag +\lambda_u^* h_u^\dag)\,,
 \label{eq:heffdstrho}
\eeq
where the second part is just the h.c. of the first part to make the whole
Hermitian, and $\lambda_c = V_{cb} V_{ud}^*$ and 
$\lambda_u = V_{ub} V_{cd}^*$ as before.
The term $\lambda_c h_c$ includes a $b \to c$ transition and 
creation of a $\bar u d$ pair, and $\lambda_c^* h_c^\dag$ includes
a $\bar b \to \bar c$ transition and 
creation of a $\bar d u$ pair, etc. Again, $h_{c,u}$ contain the effect
of strong interaction to all order.

Assuming that $CP$ violation occurs solely through the complex
$CKM$ phases, we should have
\beqa
    I\,h_c\,I^\dag = h_c^\dag\,,&\quad& I\,h_u\,I^\dag = h_u^\dag\,,\nonumber \\
    I\,h_c^\dag\,I^\dag = h_c \,,&\quad& I\,h_u^\dag\,I^\dag = h_u\,,
\eeqa
which makes $H_{\rm eff}$ invariant under $CP$ if $\lambda_{c,u}$ are real.
Then, $a_\lambda$ defined in (\ref{eq:fouramppol}) 
with $f = D^{*-}\rho^+$ can be written as
\beq
    a_\lambda \equiv \bra f_\lambda| H_{\rm eff} | B^0\ket 
    =  \bra f_\lambda| \lambda_c^* h_c^\dag | B^0\ket
    =  \lambda_c^*  \bra f_\lambda| h_c^\dag | B^0\ket\,.
\eeq
Comparing with (\ref{eq:fouramppol})  indentifies $F_{c\lambda}$ as
\beq
      F_{c\lambda} = \bra f_\lambda| h_c^\dag | B^0\ket\,.
\eeq
Similarly,
\beq
     \bar F_{c\lambda} = \bra \bar f_\lambda| h_c | \bar B^0\ket\,.
\eeq
Then, we have (again with $\eta_B=1$)
\beq
     F_{c\lambda} = 
      \underbrace{\bra f_\lambda|\,I^\dag}_{\dspl \bra\bar f_{-\lambda}|}\,
      \underbrace{I\, h_c^\dag \,I^\dag}_{\dspl h_c}\,
      \underbrace{I\, | B^0\ket}_{\dspl |\bar B^0\ket}\,
     = \bar F_{c-\lambda}\,.\quad (QED)
  \label{eq:flamqed}
\eeq
The proof of $\bar F_{u\lambda} = F_{u-\lambda}$ (helicity) proceeds the same way.

The relations (\ref{eq:dpicpfa}) is proved
similarly. Here, the effective hamiltonian is again written in the
form (\ref{eq:heffdstrho}) and is invariant under $CP$ if the $CKM$
factors are real. 
With $J=s_1=s_2=\lambda=0$ in (\ref{eq:CP2bgen}), the transformation of
the final state $f = D^{(*)-}\pi^+$ under $CP$ is
\beq
      I\,|f\ket = \eta_D\eta_\pi|\bar f\ket\,,\quad
      I\,|\bar f\ket = \eta_D^*\eta_\pi^*|f\ket\,,
\eeq
with proper choice of $CP$ phases (you can set $\lambda = 0$ in
(\ref{eq:cpflam})).
The quantities $F_{c,u}$ and $\bar F_{c,u}$ are
identified as
\beqa
   F_c = \bra f| h_c^\dag | B^0\ket\,,&\quad&
         \bar F_c = \bra \bar f| h_c | \bar B^0\ket\,, \nonumber\\
   F_u = \bra\bar f| h_u | B^0\ket\,,&\quad&
        \bar F_u = \bra f| h_u^\dag | \bar B^0\ket\,. 
\eeqa
The final result is obtained by simply setting $\lambda=0$ in 
(\ref{eq:flamqed}):
\beq
    F_{c} = \bar F_{c}\,,\quad \hbox{and similarly}\quad
    F_{u} = \bar F_{u}\,.
\eeq

\subsection{Derivation of $p/q$}

The two mass-eigenstates are the eigenvectors of the Schrodinger
equation in the $B^0$-$\bar B^0$ space:
\beq
    i{d\over dt}\Psi = H \Psi\,,
\eeq
where $\Psi$ is a two component vector 
and $H$ is a $2\times2$ matrix in the $B^0$-$\bar B^0$ space:
\beq
     \Psi = \pmatrix{a\cr b}\,,\quad
    H = M + i{\Gamma\over 2} =
    \pmatrix{ \bra B^0| H_{\rm eff} |B^0\ket & \bra B^0| H_{\rm eff} |\bar B^0\ket \cr
       \bra \bar B^0| H_{\rm eff} |B^0\ket & \bra \bar B^0| H_{\rm eff} |\bar B^0\ket
             }
\eeq
where $M$ and $\Gamma$ are hermitian matrixes. When $\gamma_a = \gamma_b$, the
decay part decouples and the mass matrix part (mixing part) only should be considered;
thus, we will drop $\Gamma$. The $CPT$ invariance allows one to write
\beq
    H = M = \pmatrix{m & \mu \cr
                    \mu^* & m}\,,\quad (M: {\rm real}\,)\,.
\eeq
In particular,
\beq
  \mu = \bra B^0| H_{\rm eff} |\bar B^0\ket \,,\quad 
  \mu^* = \bra \bar B^0| H_{\rm eff} |B^0\ket\,.
\eeq
The eigenvalues are
\beq
     \det\pmatrix{m-\lambda & \mu \cr
                    \mu^* & m-\lambda} = 0\,,\quad\to\quad
    \lambda = m \pm |\mu|\,.
\eeq
Let's define the eigenvector for the heavier of the two to be $pB^0+q\bar B^0$, which
then should satisfy
\beq
 \pmatrix{m & \mu \cr
                    \mu^* & m} \pmatrix{p\cr q} = (m+|\mu|)  \pmatrix{p\cr q} \,.
\eeq
The top component (the $B^0$ coefficient) of this equation gives
\beq
    mp + \mu q = mp + |\mu| p \quad\to\quad {p\over q} = {\mu\over|\mu|}\,.
  \label{eq:pqtemp}
\eeq

On the other hand, the $B^0\leftrightarrow \bar B^0$ transition 
is caused by the box diagram at the lowest order whose
effective  Hamiltonian can be written as
\beq
  H_{\rm eff} = (V_{tb}V_{td}^*)^2 h_{b\bar d\to d\bar b} + (h.c.)\,,
\eeq
where $h_{b\bar d\to d\bar b}$ is the effective hamitonian that transforms
$\bar B^0$ to $B^0$ and itself transforms under $CP$ as
\beq
    I\, h_{b\bar d\to d\bar b} \, I^\dag  =  h_{b\bar d\to d\bar b}^\dag \,,\quad
    I\, h_{b\bar d\to d\bar b}^\dag \, I^\dag =  h_{b\bar d\to d\bar b} \,;
\eeq
namely, $H_{\rm eff}$ is invariant under $CP$ if it were not for 
the $CKM$ phases. Then the off-diagonal elements of $H$ are related by $CP$ as
(with $\eta_B = 1$)
\beqa  
    \mu &=& \bra B^0| H_{\rm eff} |\bar B^0\ket \nonumber \\
        &=& \bra B^0|(V_{tb}V_{td}^*)^2 h_{b\bar d\to d\bar b} |\bar B^0\ket \nonumber \\
        &=& (V_{tb}V_{td}^*)^2
          \underbrace{\bra B^0| \,I^\dag}_{\dspl \bra \bar B^0|}\,
          \underbrace{I\, h_{b\bar d\to d\bar b}\,I^\dag}_{\dspl
                 h_{b\bar d\to d\bar b}^\dag}\,
          \underbrace{I\, |\bar B^0\ket}_{\dspl |B^0\ket} \nonumber \\
        &=&  {(V_{tb}V_{td}^*)^2 \over  (V_{tb}^*V_{td})^2}
       \bra\bar B^0|(V_{tb}^*V_{td})^2 h_{b\bar d\to d\bar b}^\dag |B^0\ket
        \nonumber \\
        &=&  {(V_{tb}V_{td}^*)^2 \over  (V_{tb}^*V_{td})^2}
                  \bra\bar B^0| H_{\rm eff} |B^0 \ket \nonumber \\
        &=&  {(V_{tb}V_{td}^*)^2 \over  (V_{tb}^*V_{td})^2}\, \mu^*\,.
\eeqa
Thus,
\beq
  {\mu\over\mu^*} =  {(V_{tb}V_{td}^*)^2 \over  (V_{tb}^*V_{td})^2}
   \quad\to\quad {\mu\over|\mu|} = \pm{(V_{tb}V_{td}^*)\over  (V_{tb}^*V_{td})}\,.
\eeq
Using (\ref{eq:pqtemp}), we see that $pB^0 + q\bar B^0$ with
\beq
    {p\over q} = \pm{(V_{tb}V_{td}^*)\over  (V_{tb}^*V_{td})}\,.
\eeq
This method is simple and elegant but 
cannot define the sign; in order to do so, one needs to
actually evaluate $\mu$~\cite{Branco+}:
\beq
     \mu = -{G_F^2 m_W^2\over 12\pi^2} f_B^2 m_B B_B \eta_2 S_0 
     (V_{tb}^*V_{td})^2
\eeq
where $f_B$ is the decay constant of $B^0$, $\eta_2>0$ is a QCD
correction factor, $S_0>0$ is a function of the top quark mass,
and $B_B$ is the `bag factor' of the $B$ meson which is believed to
be positive. Then, $p/q$ is now
\beq
   {p\over q} = -{V_{tb}V_{td}^*\over V_{tb}^*V_{td}}\,,
   \label{eq:povqapp}
\eeq
and these $p$ and $q$ makes $pB^0+q\bar B^0$ the heavier of the
two mass eigenstates.

In the neutral $K$ system, 
we define $p_K$ and $q_K$ in parallel to the $B$ system; namely,
the heavier ($K_L$) is defined to be $p_K K^0 + q_K \bar K^0$. Thus, $K_S$ is
\beq
     K_S = p_K K^0 - q_K \bar K^0.
\eeq
Even though there is some complication due to the
lifetime difference; the situation for the phase of $p_K/q_K$ is
essentially the same and to a good accuracy it is given by
applying $t\to c$ and $b\to s$ to (\ref{eq:povqapp}):
\beq
  {p_K\over q_K} = -{V_{cs}V_{cd}^*\over V_{cs}^*V_{cd}}\,.
   \label{eq:povqK}
\eeq
The conventions used here for $p$ and $q$ are not the same as those
in Ref.~\cite{BigiSanda}.

\end{document}